\documentclass[aoas]{imsart}

\usepackage{bm}
\usepackage{xcolor}
\usepackage{tikz, tikzscale}
\usetikzlibrary{shapes.geometric}
\usepackage{pifont}
\usetikzlibrary{fit,positioning}
\usetikzlibrary{bayesnet}
\usepackage{mathtools}
\usepackage{siunitx}
\usepackage{graphicx}
\usepackage{enumerate}
\usepackage{lscape}
\usepackage{booktabs}
\usepackage{longtable}
\usepackage{array}
\usepackage{multirow}
\usepackage{wrapfig}
\usepackage{float}
\usepackage{colortbl}
\usepackage{pdflscape}
\usepackage{tabu}
\usepackage{threeparttable}
\usepackage{threeparttablex}
\usepackage[normalem]{ulem}
\usepackage{makecell}
\usepackage{xcolor}
\newcommand{\bmI}{\ensuremath{\bm I}}

\newcommand{\bmR}{\ensuremath{\bm R}}

\newcommand{\bms}{\ensuremath{\bm s}}

\newcommand{\bmt}{\ensuremath{\bm t}}

\newcommand{\bmv}{\ensuremath{\bm v}}

\newcommand{\bmX}{\ensuremath{\bm X}}
\newcommand{\bmx}{\ensuremath{\bm x}}

\newcommand{\bmy}{\ensuremath{\bm y}}

\newcommand{\bmbeta}{\ensuremath{\bm{\beta}}}

\newcommand{\bmLambda}{\ensuremath{\bm{\Lambda}}}

\newcommand{\bmmu}{\ensuremath{\bm{\mu}}}

\newcommand{\bmTheta}{\ensuremath{\bm{\Theta}}}

\newcommand{\mT}{\ensuremath{\mathsf{T}}}

\newcommand{\bbmx}{\ensuremath{\begin{bmatrix}}}
\newcommand{\ebmx}{\ensuremath{\end{bmatrix}}}
\newcommand{\ealn}{\ensuremath{\end{aligned}}}
\newcommand{\baln}{\ensuremath{\begin{aligned}}}

\newcommand{\E}{\ensuremath{\mathrm{E}}}

\newcommand{\benum}{\begin{enumerate}}
\newcommand{\eenum}{\end{enumerate}}

\newcommand{\blonge}{\begin{longenum}}
\newcommand{\elonge}{\end{longenum}}

\RequirePackage{amsthm,amsmath,amsfonts,amssymb}
\RequirePackage[authoryear]{natbib}
\RequirePackage[colorlinks,citecolor=blue,urlcolor=blue]{hyperref}%% uncomment this for coloring bibliography citations and linked URLs
\RequirePackage{graphicx}%% uncomment this for including figures

\startlocaldefs
\theoremstyle{plain}

\newtheorem{theorem}{Theorem}[section]
\theoremstyle{definition}

\newtheorem{condition}{Condition}

\endlocaldefs

\begin{document}

\begin{frontmatter}
\title{A Bayesian Record Linkage Approach to Applications in Tree
Demography Using Overlapping LiDAR Scans}

\runtitle{A Record Linkage Approach to Tree Demography}

\begin{aug}
%%%%%%%%%%%%%%%%%%%%%%%%%%%%%%%%%%%%%%%%%%%%%%%
%% Only one address is permitted per author. %%
%% Only division, organization and e-mail is %%
%% included in the address.                  %%
%% Additional information can be included in %%
\author[A]{\fnms{Lane}~\snm{Drew}\ead[label=e1]{lane.drew@colostate.edu}\orcid{0009-0006-5427-4092}},
\author[A]{\fnms{Andee}~\snm{Kaplan}\ead[label=e2]{andee.kaplan@colostate.edu}\orcid{0000-0002-2940-889X}}
\and
\author[B]{\fnms{Ian}~\snm{Breckheimer}\ead[label=e3]{ikb@rmbl.org}\orcid{0000-0002-4698-977X}}

%%%%%%%%%%%%%%%%%%%%%%%%%%%%%%%%%%%%%%%%%%%%%%
%% Addresses                                %%
%%%%%%%%%%%%%%%%%%%%%%%%%%%%%%%%%%%%%%%%%%%%%%

\address[A]{Department of Statistics,
Colorado State University\printead[presep={,\ }]{e1,e2}}

\address[B]{Rocky Mountain Biological Laboratory\printead[presep={,\ }]{e3}}
% \address[C]{Clark Family School of Environment and Sustainability, Western Colorado University\printead[presep={,\ }]{e3}}

\end{aug}

\begin{abstract}
In the information age, it has become increasingly common for data
containing records about overlapping individuals to be distributed
across multiple sources, making it necessary to identify which records
refer to the same individual. The goal of record linkage is to estimate
this unknown structure in the absence of a unique identifiable
attribute. We introduce a Bayesian hierarchical record linkage model for
spatial location data motivated by the estimation of individual specific
growth-size curves for conifer species using data derived from
overlapping LiDAR scans. Annual tree growth may be estimated dependent
upon correctly identifying unique individuals across scans in the
presence of noise. We formalize a two-stage modeling framework,
connecting the record linkage model and a flexible downstream individual
tree growth model, that provides robust uncertainty quantification and
propagation through both stages of the modeling pipeline via an
extension of the linkage-averaging approach of
\cite{sadinleBayesianPropagationRecord2018a}. In this paper, we discuss
the two-stage model formulation, outline the computational strategies
required to achieve scalability, assess the model performance on
simulated data, and fit the model to a bi-temporal dataset derived from
LiDAR scans of the Upper Gunnison Watershed provided by the Rocky
Mountain Biological Laboratory to assess the impact of key topographic
covariates on the growth behavior of conifer species in the Southern
Rocky Mountains (USA).
\end{abstract}


\end{frontmatter}

%%%%%%%%%%%%%%%%%%%%%%%%%%%%%%%%%%%%%%%%%%%%%%
%%%% Main text entry area:

\section{Introduction}\label{section1}

The characterization and quantification of forest dynamics has been an
area of interest for ecologists for more than a century and has become
an increasingly important metric for understanding the effects of
climate change (\cite{hyyppaReviewMethodsSmallfootprint2008}).
Historical investigations of forest dynamics have relied on field
surveys over limited spatial domains, which are generally time consuming
and potentially difficult to perform
(\cite{saatchiBenchmarkMapForest2011, wenselTreeHeightDiameter1987}).
The advent and ongoing refinement of aerial laser scanning (ALS)
technology has ushered in a new age of data collection in terms of
scalability. The use of ALS data in the modeling of forest structure has
become a standard approach, as it enables researchers to examine the
health and behavior of forests at larger scales than has previously been
possible by field survey
(\cite{babcockModelingForestBiomass2016, dalponteTreecentricMappingForest2016}).
The obvious extension of these efforts is to functions that rely on
repeat measurements over time. Despite improvements in the accuracy of
ALS technology, there remains inherent uncertainty in both the scanning
mechanism and subsequent post-processing of the data as discussed by
\cite{huoIndividualTreeDetection2020}. In the existing literature, the
mechanisms employed for identifying the unique individuals from scans
across multiple time points have been largely heuristic and rely upon
manual verification, as in \cite{maQuantifyingIndividualTree2018}, and
employ a two-stage modeling schema which fails to incorporate the
uncertainty in the segmentation and matching procedures into the
downstream task. To address these issues, we present an alternative
two-stage framework utilizing a record linkage approach for spatial
location data that is capable of efficiently identifying unique
individuals across larger spatial domains while providing robust
uncertainty quantification for the linkage that may then be propagated
into the downstream modeling objective.

As our ability to collect and store data has exploded, so too has our
need to engage in record linkage (also called deduplication or entity
resolution). At its core, the field of record linkage is concerned with
the resolution of unique records across overlapping files in the absence
of a unique identifier. In this paper we use the term record linkage to
encompass the process of identifying coreferent records both between and
within files. Historically, files have represented repeated surveys over
time (\cite{steortsEntityResolutionEmpirically2015a}), or non-temporally
linked overlapping databases such as patient records across different
providers in the healthcare system
(\cite{padmanabhanApproachRecordLinkage2019}). The earliest approaches
to probabilistic record linkage, as formalized by
\cite{fellegiTheoryRecordLinkage1969}, performed probabilistic matching
between pairs of records according to a decision theoretic framework.
The field has developed consistently since its inception, and advances
in Bayesian computational methods have given rise to a new class of
probabilistic modeling approaches over the last 20 years as discussed by
\cite{Liseo2011advances}. While methods that perform matching between
pairs of records directly have remained useful (\cite{Sadinle2017}),
models built upon latent clustering structures have become increasingly
popular alongside methods addressing alternative types of data
(\cite{steortsBayesianApproachGraphical2016, liseoBayesianEstimationPopulation2011a}).
Recently, record linkage has been successfully used to improve wildlife
population inference from a series of sequential aerial photographs
(\cite{Lu2022}). In a similar vein, we introduce a record linkage model
for multi-temporal spatial location data derived from light detection
and ranging (LiDAR) scans intended to improve tree demography inference.

While record linkage is an interesting and challenging endeavor on its
own, it generally functions as the first step in the sequence of a
statistical pipeline as discussed by
\cite{kaplanPracticalApproachProper2022}. We implement our spatial
record linkage model in a two-stage framework in which the record
linkage and downstream task are performed sequentially. Our proposed
approach performs the downstream modeling task using a randomly sampled
subset of iterations from the posterior linkage structure, which
propagates the uncertainty from the linkage into the second stage of the
pipeline according to the linkage-averaging (LA) approach of
\cite{sadinleBayesianPropagationRecord2018a}. Crucially, in the LA
framework, the linkage is not informed by the downstream task.
Consequently, the output from the first stage record linkage model may
be used as the input for a variety of downstream models, offering
researchers a high degree of flexibility when adopting this modeling
framework.

In our application, we pair the spatial record linkage model with an
individual tree growth model, where we define growth as the change in
canopy volume between surveys on an annual scale. The empirical data is
comprised of LiDAR scans of Gunnison National Forest from 2015 and 2019
provided by the Rocky Mountain Biological Laboratory (RMBL). The
individual tree crown polygons and associated attributes were obtained
from a 1/3 \unit{\meter} resolution canopy height model using the
\texttt{ITCsegment} (\cite{dalponteTreecentricMappingForest2016})
algorithm in the \texttt{lidR} (version 3.4,
\cite{rousselLidRPackageAnalysis2020}) package in R. We note the record
linkage and growth models are both designed to account for various
sources of biological variation and measurement error in the data
collection and post-processing procedures.

The remaining structure of the paper is as follows. In Section
\ref{section2}, we highlight the ecological hypotheses and empirical
data that motivate our novel modeling approach. In Section
\ref{section3}, we introduce the relevant notation for the spatial
record linkage model and downstream growth model. Additionally, we
outline the computational strategies used to fit the model. In Section
\ref{section4}, we provide a discussion of the theoretical justification
for the LA approach in a general auxiliary data task setting. In Section
\ref{section5}, we provide an analysis of the empirical data and in
Section \ref{section6}, we examine the performance of the proposed
modeling approach in a series of numerical experiments on simulated
data. We conclude with a discussion and directions for future work in
Section \ref{section7}.

\section{LiDAR Derived Individual Tree Characteristics from Multi-temporal Scans of Snodgrass Mountain}\label{section2}

We apply our two-stage modeling approach to identify unique trees across
time points and to estimate spatial patterns of tree growth as related
to certain environmental drivers in a spruce-fir forest site located in
the Southern Rocky Mountains (USA). The study site is a two square
kilometer forested domain located on Snodgrass Mountain near the site of
RMBL in the vicinity of Crested Butte, Colorado. The domain spans
montane to lower subalpine mountain slopes at elevations from
2891-3395\unit{\meter} and experiences a cold continental climate with
persistent seasonal snowpack accounting for the majority of annual
precipitation (\cite{carroll_efficiency_2020}). Evergreen forests in the
domain are dominated by Engelmann spruce (\emph{Picea engelmannii}) and
subalpine fir (\emph{Abies lasiocarpa}), which account for more than
80\% of the tree canopy. These forests also contain scattered lodgepole
pine (\emph{Pinus contorta}) and Rocky Mountain Douglas-fir
(\emph{Pseudotsuga menziesii subsp. glauca}). Deciduous quaking aspen
(\emph{Populus tremuloides}) forms large single-species stands on lower
slopes of the study area, but these areas were excluded from the
analysis due to the difficulty of assessing the growth of this species.

Forest structural data was collected for the study area via LiDAR in two
intervals during 2015 and 2019. Both scans were collected from an
airplane-based sensor in late summer before the drop of deciduous
leaves. The laser scanner records discrete peaks of reflected energy at
near-infrared wavelengths and uses the integrated Real-time Kinematic
(RTK) sensor position and estimated time-of-flight of laser pulses to
locate laser reflections (``returns'') in geographic space. The scanning
process yields a dense (8-16 pts / \unit{\meter^2}) cloud of
3-dimensionally located returns representing reflections from the
ground, tree canopies, and other reflective surfaces. Details of each
LiDAR dataset are provided in Table \ref{scanattributes}.

\begin{table*}
\caption{LiDAR data collection attributes for the 2015 and 2019 scans provided by RMBL.}
\label{scanattributes}
\begin{tabular}{@{}lll@{}}
\hline
Scan Attribute & 2015 Scan & 2019 Scan \\
\hline
Acquisition Dates & August 7, 2015 and August 10, 2015 & August 21 – September 24, 2019\\
Aircraft Used & Piper Navajo & Cessna Caravan \\
Sensor &    Reigl (Leica) Q1560 &   Riegl (Leica) VQ1560i \\
Maximum Returns / Pulse & 5 &   15 \\
Target Pulse Density & Average 8 pulses $\unit{\meter^2}$ & Average 2 pulses $\unit{\meter^2}$ \\
Realized Point Density &        9.4 pts / $\unit{\meter^2}$ & 9.4 pts / $\unit{\meter^2}$ \\
Survey Altitude (AGL) & \qty[mode = text]{550}{\unit{\meter}} & \qty[mode = text]{1159}{\unit{\meter}} \\
Field of View & \ang[mode = text]{58} & \ang[mode = text]{58.5}\\
\hline
\end{tabular}
\end{table*}

The LiDAR-derived point clouds were segmented and summarized to yield
estimates of per-tree structural characteristics including tree top
locations, maximum heights, and canopy volumes using functions in the R
package \texttt{lidR} (version 3.4,
\cite{rousselLidRPackageAnalysis2020}). Although numerous approaches
exist to segment individual trees in LiDAR data (see
\cite{aubry-kientz_comparative_2019} for a recent comparison), for this
analysis we adopted the commonly-used \texttt{ITCsegment} algorithm
(\cite{dalponteTreecentricMappingForest2016}). \texttt{ITCsegment} is a
region-growing approach which iteratively incorporates points into
candidate tree canopies starting at a set of seed locations. Seed
locations (putative tree tops) were selected using a local maximum
filter, identifying laser returns with high heights relative to a
height-dependent local neighborhood. Canopy volumes were calculated by
summing canopy heights for each segment using a
\qty[mode = text]{1}{\unit{\meter}} resolution canopy surface model
generated using the \texttt{pit-free} algorithm implemented in
\texttt{lidR}. The segmentation and canopy surface model generation
steps generate imperfect representations of individual tree locations
and crown geometries. Errors in sensor geo-positioning and ranging
measurements can lead to systematic spatial shifts in datasets collected
at different times. Moreover, the location and trajectory of individual
laser pulses differs between scans, leading to small amounts of
variability in estimated maximum heights and crown locations as
discussed by \cite{poorazimyFeasibilityBiTemporalAirborne2022}.

\begin{figure}
\centering
\includegraphics[width = 400pt]{"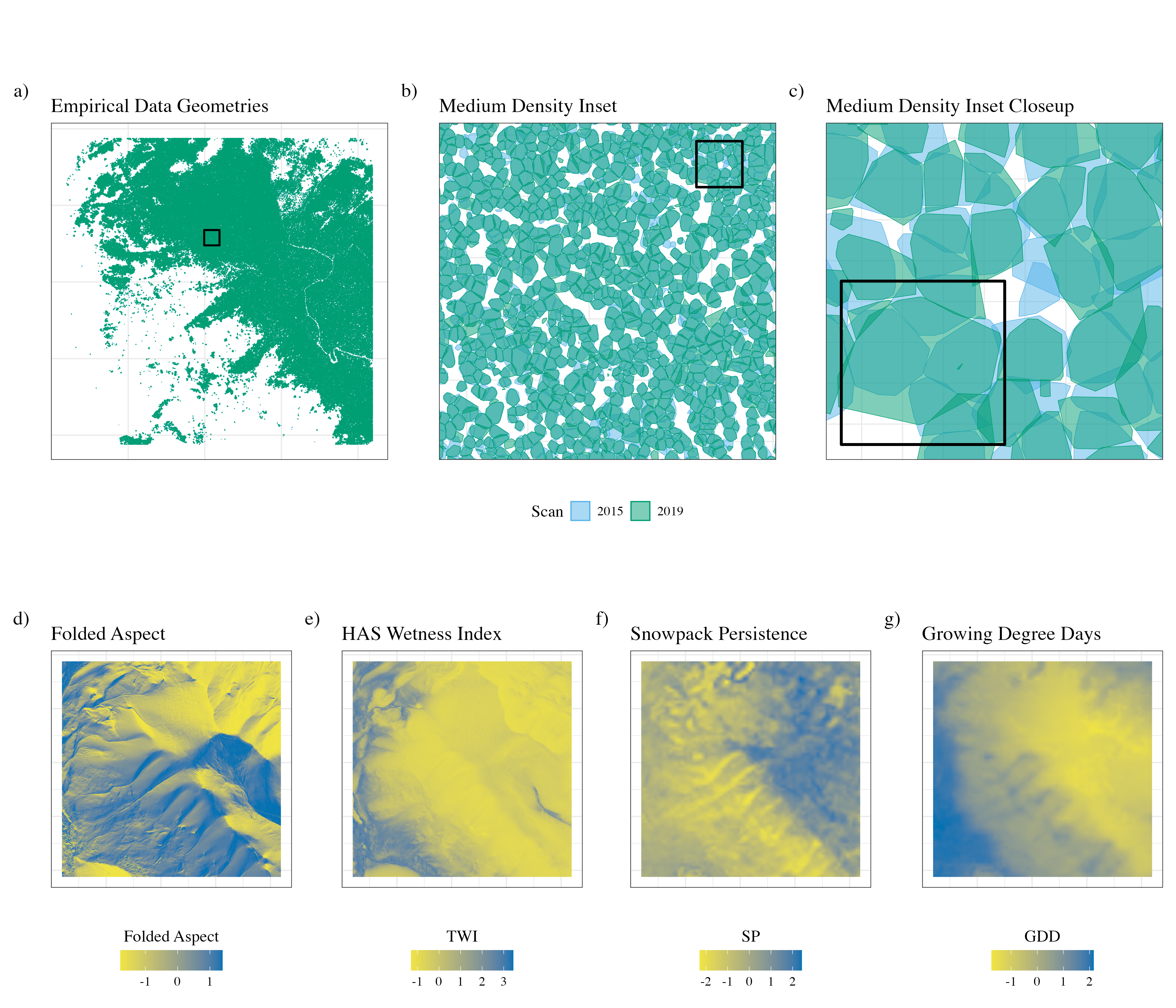"}
\caption{Plots a) and b) show the derived crown geometries from the 2015 and 2019 LiDAR scans performed by RMBL for the full datasets and a medium density inset (outlined in plot a)). Plot c) shows a closeup from the inset in b), highlighting an instance in which multiple trees in the first file overlap with a single tree in the second file. Plots d) - g) show the scaled raster images for the topographic covariates of interest over the study domain.}
\label{emp_rast_plot}
\end{figure}

In Western North American conifer forests, tree growth is thought to be
constrained by the (potentially interacting) availability of water and
energy
(\cite{buechling_climate_2017, heilmanEcologicalForecastingTree2022}).
To investigate these constraints, we assembled estimates of key water
and energy proxies across the domain from diverse remote sensing
datasets. A \qty[mode = text]{1}{\unit{\meter}} resolution LiDAR-derived
Digital Elevation Model (\cite{goulden_neon_2020}) was used to compute
topographic aspect ``folded'' about the north-south axis to distinguish
high solar-radiation south-facing slopes from low solar-radiation
north-facing slopes. We also computed a topographic wetness index (TWI)
(\cite{nobreHeightNearestDrainage2011}) as a water availability proxy.
We augmented these topographic proxies with gridded climate data
interpolated from weather station and microclimate sensors as well as
satellite-derived maps of the persistence of seasonal snowpack
(\cite{breckheimer_integrated_2023}). Data for snowpack persistence and
growing degree days was aggregated annually from 2015 to 2019 and the
median observed values were used during modeling to capture the relative
impact of these covariates over the period between LiDAR scans. Details
pertaining to the covariates may be found in Table \ref{covariateinfo}
along with a visualization of the derived tree geometries and raster
images in Figure \ref{emp_rast_plot}.

\begin{table*}
\caption{Designations and details for the topographic covariates included in the analysis.}
\label{covariateinfo}
\begin{tabular}{@{}llll@{}}
\hline
Growth Constraint & Covariate & Data Source &   Resolution (\unit{\meter}) \\
\hline
Energy & Folded Aspect & \cite{goulden_neon_2020} & 1 \\
Energy & Growing Degree Days & \cite{breckheimer_integrated_2023} & 30 \\
Water & HAS Wetness Index & \cite{goulden_neon_2020, nobreHeightNearestDrainage2011} & 1 \\
Water & Snowpack Persistence & \cite{breckheimer_integrated_2023} & 30 \\
\hline
\end{tabular}
\end{table*}

\section{Models and Notation}\label{section3}

In this section we detail the spatial record linkage and growth models
employed in our two-stage LA approach. We first define the necessary
notation and present the proposed spatial record linkage model before
developing the downstream individual tree growth model. We note that
throughout this section, distributions identified with subscripts refer
to truncated distributions over the specified bounds. For example a
truncated Inverse-Gamma distribution with parameters \(c\) and \(d\)
over the range \([0, b]\) is denoted as
\(\text{Inverse-Gamma}_{[0,b]}(c,d)\). We finish this section with a
discussion of the computational strategies employed to facilitate
scalability of the record linkage model to spatial domain sizes that are
of practical interest.

\subsection{Record Linkage Model}\label{section3.1}

We begin by presenting the spatial record linkage model as a standalone
component to introduce the model structure and to establish a baseline
for inclusion in modeling pipelines with alternative downstream tasks.
We provide a general model capable of handling two files, which is also
capable of performing deduplication within files. We employ a Bayesian
hierarchical structure based on latent matching, as discussed by
\cite{steortsBayesianApproachGraphical2016} and
\cite{liseoBayesianEstimationPopulation2011a}, such that records are
linked to unobserved latent entities with true field values instead of
being probabilistically matched to other records directly through
comparison vectors, as in the work of
\cite{fellegiTheoryRecordLinkage1969} and \cite{Sadinle2017}. In the
spatial record linkage model, the value associated with the latent
entity is the true unobserved location of the individual. We treat the
observed data (i.e.~location) as a noisy observation of the latent
location and include provisions for different potential sources of
noise. We consider error introduced as a function of translation and
rotation in the data collection process, post-processing of the data,
and due to biological mechanisms. We specify the data model and relevant
notation as follows.

The model is constructed to handle two files, where the files are
indexed by \(i = 1,2\) with relative size \(n_i\) for each file. The
records within files are indexed from \(j = 1, \dots, n_i\) such that
the total number of observed records is \(n = \sum_{i=1}^2 n_i\). We
denote the observed location data for the \(j^{th}\) record in file
\(i\) as \(\bmy_{ij}\), where \(\bmy_{ij}\) is a numerical vector of
length 2, i.e.~\(\bmy_{ij}=\left(x,y \right)_{ij}\), corresponding to
the spatial coordinates of the record in the
\(\left(x,y \right)\)-plane. We note that in our application the term
file is synonymous with a LiDAR scan and record with an individual
identified tree such that \(n\) is the total number of individual trees
detected across all scans.

We define the latent location vector as \(\bms_{j'}\), where
\(j'=1,\dots,N\) such that \(N\) is the maximum number of unique latent
individuals in the population under consideration. The observed
locations \(\bmy_{ij}\) are modeled as noisy versions of the latent
locations \(\bms_{j'}\). We assume that the record set \(\bmy_{1j}\)
exists in the same space as the latent locations, while the record set
\(\bmy_{2j}\) is modeled as a transformed version of the associated
\(\bms_{j'}\), where we restrict the possible transformations considered
to rotation and translation.

The linkage structure, which identifies the relationship between the
observed records and the latents, is a vector of length \(n\) denoted as
\(\bmLambda = \left\{\lambda_{ij}: i=1,2; \ j=1,\dots,n_i \right\}\),
where \(\lambda_{ij}\) is an integer indicating which \(\bms_{j'}\) the
\(j^{th}\) record in file \(i\) refers to. The linkage is implicitly
dependent on the maximum latent population size \(N\), as
\(\lambda_{ij} \in \left\{1, \dots, N \right\}\). The specified linkage
structure naturally defines a set of \(N\) clusters denoted
\(\mathcal{C}(\bmLambda)\), which specify the records that are linked to
each \(\bms_{j'}\) such that
\(\mathcal{C}\left(\bmLambda\right) = \left\{ \mathcal{C}_1,\dots,\mathcal{C}_N \right\}\).
The individual clusters are defined as the sets
\(\mathcal{C}_{j'} = \left\{ (i,j): \lambda_{ij} = j' \right\}\) for
\(j' = 1, \dots, N\), and we note that the clusters may be empty or may
contain records from the same file in addition to records across files
highlighting the capacity of the model for performing record linkage and
deduplication simultaneously. In the context of our application,
duplicates within files potentially occur during the LiDAR processing
due to the segmentation algorithm, such that a single individual may be
erroneously split into multiple entities as seen in panel c) of Figure
\ref{emp_rast_plot}.

Previous record linkage modeling approaches are known to be sensitive to
the specification of \(N\), which functions as a hyperparameter in the
model and quantifies our belief regarding the upper bound on the number
of unique entities across all files
(\cite{steortsBayesianApproachGraphical2016}). We specify
\(N=q \times\max\left(n_i\right)\) where the scale factor \(q\) is
chosen to reflect the assumed degree of overlap between the files and
such that \(N \leq n\). Alternatively, a practitioner could specify
\(N=n\) to avoid making any a priori assumptions about the number of
unique individuals across files. Although not explicitly a parameter in
the model, we do effectively obtain an estimate of the number of unique
individuals across files which is often of interest in studies
investigating species abundance and provides some sense of the effective
sample size for estimation of the downstream model parameters.
Additional discussion regarding the specification of \(N\) is provided
in the Supplementary Materials in Appendix A
(\cite{drewBayesianRecordLinkage2025sup}).

The data model, which describes the relationship between the observed
point patterns and \(\bms\), allows us to model the variation produced
by the underlying biological process (assumed to be tree growth in this
application) separately from the error introduced in the LiDAR scanning
and post-processing procedures by incorporating the image alignment
framework introduced by \cite{greenBayesianAlignmentUsing2006}. The
model is specified as follows, \[
\bmy_{ij} \mid \bms_{\lambda_{ij}}, \sigma^2, \bmt_i, \theta_i, D \sim \text{Normal}_{2,[D]} \left(\bmR\left(\theta_i\right)\left(\bms_{\lambda_{ij}} - \bmmu_D\right) + \bmt_i + \bmmu_D, \sigma^2 \bmI \right),
\] for \(i = 1,2\). The rotation, \(\bmR(\theta_i)\), is the standard
counterclockwise rotation matrix given by \[
\bmR\left(\theta_i\right) = \bbmx \cos\left(\theta_i\right) & -\sin\left(\theta_i\right) \\ \sin\left(\theta_i\right) & \cos\left(\theta_i\right) \ebmx
\] and \(\bmt_i\) is the two dimensional translation vector. We allow
the rotation angle and translation to vary across files (i.e.~scans).
The rotation for each file is around the midpoint, denoted \(\bmmu_D\),
of the spatial domain of interest \(D\). We also note that the record
set \(\bmy_{1j}\) can be expressed in terms of the rotation and
translation framework with fixed \(\theta_1 = 0\) and
\(\bmt_1 = \bbmx 0 & 0 \ebmx^\mT\), assuming that the records in the
first file exist in the same space as the latent locations, \(\bm{s}\),
to reduce the effective number of parameters in the model. We adopt this
expression to simplify notation going forward. The full spatial record
linkage model is specified as follows \[
\baln
\bmy_{ij} \mid \bms_{\lambda_{ij}}, \sigma^2, \theta_i, \bmt_i, D &\stackrel{ind}{\sim} \text{Normal}_{2,[D]}\left(\bmR\left(\theta_i\right)\left(\bms_{\lambda_{ij}} - \bmmu_D\right) + \bmt_i + \bmmu_D, \sigma^2 \bmI \right) \\
\bms_{j'} \mid N &\stackrel{iid}{\sim} \text{Uniform}\left(D^*\right) \\
\sigma^2 &\sim \text{Inverse-Gamma}_{[0, b_\sigma]} \left(c_\sigma, d_\sigma\right) \\
\lambda_{ij} \mid N &\stackrel{iid}{\sim} \text{Uniform}\left\{1,\dots,N\right\} \\
\theta_i &\propto \text{exp} \left(\kappa \cos\left(\nu\right) \cos\left(\theta_i\right) + \kappa \sin\left(\nu\right) \sin\left(\theta_i\right)\right) I\left\{\left|\theta_i\right| < b_\theta \right\} \\
\bmt_i &\sim \text{Normal}_2 \left(\bm0, \sigma_t^2 \bmI\right),
\ealn
\] where the prior for \(\theta_i\), the rotation parameter for file
\(i\), is the kernel of a truncated von Mises distribution, as discussed
by \cite{greenBayesianAlignmentUsing2006}.

As mentioned above, we model the observed locations as noisy
transformations of the unobserved true \(\bms_{j'}\) according to a
Gaussian noise process. We specify the underlying latent point process,
\(\bms\), to follow a uniform distribution over a slightly expanded
spatial domain \(D^*\), such that \(D \subseteq D^*\), to allow for the
possibility that the true location of an individual is outside of the
observed spatial domain (i.e.~the tree base is located outside of \(D\),
but the observed tree crowns are inside). We assume the simplest and
least informative prior specification for \(\bms\), corresponding to
complete spatial randomness with a fixed number of points. This prior
allows the observed locations to provide the bulk of the information in
determining the distribution of the \(\bms_{j'}\) and enables this
distribution to vary adequately over space. In practice, users could
specify more complicated latent process models, which may fit the
observed data more precisely but with additional assumptions,
computational cost, and complexity (see
\cite{leiningerBayesianAnalysisSpatial2014a} for an in-depth discussion
of point process models in a Bayesian hierarchical framework). We
provide the derivations of the joint posterior and full conditional
distributions for the model alongside the algorithm for fitting the
model in Appendix A of the Supplementary Materials
(\cite{drewBayesianRecordLinkage2025sup}).

Many latent record linkage modeling approaches employ a hit-miss
mechanism for whether the observed records are a noisy distortion of the
true latent value in a given field as in
\cite{steortsBayesianApproachGraphical2016}. In contrast, our model
treats every observed record as a noisy copy of the latent as we are
dealing with spatial locations over a continuous domain. In the context
of our motivating application, the observed locations are the tree
crowns while the latent location is the tree base. Thus, the assumption
that every observed location is a noisy observation of the truth has a
clear physical interpretation in this case as well. We place a conjugate
truncated Inverse-Gamma distribution prior on the measurement error
parameter \(\sigma^2\), where the upper bound corresponds to the maximum
displacement that would be considered plausible based on the biological
mechanisms of tree growth and the calibration of the LiDAR scanning
equipment.

We note that our record linkage inspired modeling approach could be
interpreted as a microclustering approach such that cluster sizes remain
small even as the number of records grows.
\cite{betancourtRandomPartitionModels2022} introduced a class of random
partition models with microclustering behavior to achieve a similar
goal, i.e.~guaranteeing that clusters remain small by assuming
exchangeable sequences of clusters instead of exchangeable data points.
In contrast to this approach, we obtain microclustering behavior in our
model through the specification of \(N\) and the prior on \(\sigma^2\)
by guaranteeing a maximum number of unique latent locations and distance
that observed locations may be from their associated true latent
location, despite the fact that the prior for the linkage does not
explicitly guarantee this property.

\subsection{Downstream Growth Model}\label{section3.2}

We now turn to the individual tree growth model we employ in this
application. The model leverages the known allometric relationship
between size and growth by using a flexible nonlinear function of the
generalized Michaelis-Menten type to describe the annual individual
growth-size curve while allowing for measurement error in the observed
growth (measured by the change in canopy volume). Michaelis-Menten
functions have been applied broadly across biological and ecological
growth models, as described by
\cite{lopezGeneralizedMichaelisMentenEquation2000a} and
\cite{bolkerEcologicalModelsData2008}. The generalized Michaelis-Menten
function can take on a range of shapes from a logistic to a sigmoidal
curve depending on the parameterization, where our specification may be
seen in Eq. (\ref{eq1}) below. The relative simplicity and flexibility
of the function class combined with the clear biological interpretations
of the parameters make this model a compelling choice. Michaelis-Menten
growth models have been found to be ideal for describing the
relationship between diameter at breast height (DBH) and height for
trees (\cite{Barbosa2019, Brahma2017}), which is analogous to the
size-growth relationship between canopy volume and tree growth in our
application. Although literature employing a Michaelis-Menten function
in a measurement error model is scarce, the extension is natural and
straight forward.

Our specification of the growth function incorporates topographically
derived covariates, discussed in Section \ref{section2}, allowing us to
better understand the impact environmental drivers have on the growth of
conifer species. We introduce relevant notation for the downstream
growth model and clarify the relationship with the spatial record
linkage model as follows.

We previously defined the set of clusters
\(\mathcal{C}\left(\bmLambda\right)\) derived from the linkage structure
of the spatial record linkage model, and we further restrict this set to
the clusters for which growth is observed. We define the set of growth
clusters \(\mathcal{C}^G\left(\bmLambda\right)\), with respect to
several ecological conditions, which correspond to the individual trees
identified by the linkage model for which a plausible change in canopy
volume has occurred between the two time points. For notational clarity,
we introduce the functions \(\min^f\left(\right)\) and
\(\max^f\left(\right)\) which return the minimum and maximum file index,
respectively, for a given cluster. We note the implicit ordering in the
file indices relative to time, such that the first file is the oldest
and the second file is the most recent. We denote the observed canopy
volume of the \(j^{th}\) record in file \(i\) as \(v_{ij}\), measured in
cubic meters, where the file index is synonymous with the data
collection time point associated with the file. The set of growth
clusters may be defined accordingly as
\(\mathcal{C}^G\left(\bmLambda\right) = \{ \mathcal{C}_{j'}: \max^f\left(\mathcal{C}_{j'}\right) \neq \min^f\left(\mathcal{C}_{j'}\right) \  \& \ r_1\cdot v^*_{\min^f\left(\mathcal{C}_{j'}\right)} < v^*_{\max^f\left(\mathcal{C}_{j'}\right)} < r_2\cdot v^*_{\min^f\left(\mathcal{C}_{j'}\right)} \}\)
such that
\(\mathcal{C}^G\left(\bmLambda\right) \subseteq \mathcal{C}\left(\bmLambda\right)\).
Where \(v^*_{\min^f\left(\mathcal{C}_{j'}\right)}\) and
\(v^*_{\max^f\left(\mathcal{C}_{j'}\right)}\) denote the summed volumes
of the records associated with the minimum and maximum file indices in
the cluster. This implicitly defines a procedure that merges the volumes
of linked records within files as a result of deduplication from the
linkage model. The hyperparameters \(r_1\) and \(r_2\) control the lower
and upper bounds for the change in canopy volume for a growth cluster
relative to the typical and biologically feasible growth behavior for
the time interval between the observed records and such that
\(0 < r_1 < r_2\). For example, if we specify \(r_1 = .9\) and
\(r_2 = 1.6\), then we would restrict our set of growth clusters to
those that saw between a 10\% loss and a 60\% increase in canopy volume
over the interval between measurements. We exclude the clusters which do
not satisfy the specified growth rate constraints from the set of growth
clusters used to estimate the growth model parameters. We note that the
excluded clusters from the linkage model correspond to changes in canopy
volume due to abiotic factors or obvious errors in the linkage resulting
in biologically implausible growth rates. While it is possible that a
tree may experience a decline in canopy volume over time (e.g.~during
the mortality process), we have chosen to emphasize a method that
focuses on the growth of healthy trees with the recognition that this
restricts our understanding of the growth relationship conditional on
the fact that a tree grew, or experienced a small enough decline in
canopy volume that the change could be attributed to errors in the LiDAR
scanning and post processing.

The row vector \(\bmx_{s_c}\), of length \(p + 1\), contains \(p\)
observed covariates at the latent location \(\bms_c\) for the growth
cluster \(\mathcal{C}^G_c\) with first element corresponding to the
baseline growth rate asymptote. We note that topographic covariates
(Folded Aspect, Growing Degree Days, HAS Wetness Index, and Snowpack
Persistence in this application) are assumed to be centered and scaled
across the entire surface of the domain of interest \(D\) prior to
inclusion in the model.

For each growth cluster
\(\mathcal{C}^G_c \in \mathcal{C}^G\left(\bmLambda\right)\), \(g_{c}\)
is the observed annual growth for cluster \(c\) and is defined as a
function of the first and last cumulative volume measurements for the
linked record set \(\mathcal{C}^G_c\) such that \[
g_{c} = \frac{v^*_{\max^f\left(c\right)} - v^*_{\min^f\left(c\right)}}{t \left( \max^f \left(c\right)\right) - t\left( \min^f \left(c \right) \right)}.
\] The function \(t()\) returns the year associated with the file index
so that the difference in observed canopy volumes is scaled by the
length of the interval between measurements to place the observed growth
on an annual scale. We model the observed annual growth as a function of
the true growth, which we specify as a Michaelis-Menten type function
dependent upon the initial observed canopy volume and the environmental
covariates associated with the record's latent location \(\bms_c\),
while allowing for measurement error, such that
\begin{equation}\label{eq1}
g_{c} \mid \gamma, \bmbeta, \tau, \bmLambda, \bmx_{\bms_c}, \bmv^* \sim \text{Skewed t} \left( \mu_c, \tau, \delta, \omega \right), \text{ for } \mu_c = \frac{\left(\bmx_{s_c} \bmbeta\right) {v^*_{\min^f\left(c\right)}}^\alpha}{\gamma^\alpha + {v^*_{\min^f\left(c\right)}}^\alpha}.
\end{equation} As mentioned above, a notable advantage of the
generalized Michaelis-Menten style growth function is that the
parameters of the true growth function have clear biological
interpretations. The linear component of the function adjusts the
maximum growth asymptote as a function of the covariates at a given
location. The parameter \(\gamma\) controls the size at which the growth
rate saturates due to size scaling, establishing the inflection point of
the growth curve. The parameter \(\alpha\) controls the curvature of the
growth function, where values of \(\alpha > 1\) result in a sigmoidal
curve and values of \(\alpha \leq 1\) result in a shape more akin to a
logistic curve. While the true growth is generally assumed to be
non-negative, our model allows for the observed growth to be negative as
a function of measurement error.

The full growth model is defined as follows \[
\baln
g_c \mid \gamma, \bmbeta, \tau, \bmLambda, \bmx_{\bms_c}, \bmv^* &\stackrel{ind}{\sim} \text{Skewed t} \left( \mu_c, \tau, \delta, \omega \right) \\
\tau &\sim \text{Uniform} \left(0, b_\tau \right) \\
\delta &\sim \text{Normal}_{[-1,1]}\left( 0, \sigma^2_\delta \right) \\
\omega &\sim \text{Gamma} \left(2, b_\omega \right) \\
\gamma &\sim \text{Uniform}\left(a_\gamma, b_\gamma \right) \\
\alpha &\sim \text{Beta}_{[c_\alpha, d_\alpha]} \left(a_\alpha, b_\alpha \right) \\
\beta_0 &\sim \text{Normal} \left(\mu_0, \sigma^2_0 \right) \\
\beta_k &\stackrel{ind}{\sim} \text{Normal} \left(\mu_{\beta_k}, \sigma_{\beta_k}^2 \right), \text{ for $k = 1, \dots, p$},
\ealn
\] where the specification of the hyperparameters is informed by the
advice of domain science experts, while being adequately diffuse where
appropriate. We model the observed growth, annual change in canopy
volume, as a nonlinear function of the initial size and a set of
environmental covariates at the latent location \(\bms_c\) of the
individual with skewed measurement error derived from the LiDAR
processing algorithm described in Section \ref{section2}. We place a
weak Uniform prior on \(\gamma\) with the lower and upper bounds
specified as the minimum reasonable growth saturation value as a
function of size and the maximum size observed in the first file, as the
parameter corresponds to the size at which the growth rate reaches its
half maximum. We specify a shifted and scaled Beta prior for the shape
parameter \(\alpha\), which controls the curvature of the growth
function, where the specified bounds limit the degree of possible
curvature. A non-informative version of this prior takes
\(a_\alpha = b_\alpha = 1\) corresponding to a Uniform distribution over
the specified range. We assume the individual \(\beta_k\) coefficients
are independent a priori and assign appropriately diffuse Normal priors
with the understanding that all covariates have been centered and scaled
prior to inclusion in the model. We place a Gamma prior on \(\omega\),
where \(\omega\) controls the kurtosis of the distribution. Finally, we
specify a truncated Normal prior for the skewness parameter \(\delta\),
where the truncation bounds follow the support of the parameter. In this
model formulation we adopt a nonlinear regression skew-t error model, as
presented by \cite{delacruzBayesianAnalysisNonlinear2009}, to account
for the observed structure of our empirical data. For our specific
skew-t density, we follow the formulation of
\cite{hansenAutoregressiveConditionalDensity1994} and perform the
appropriate location and scale adjustments such that \(\mu_c\) and
\(\tau\) are the mean and variance of the distribution respectively when
fitting the model. We do note however, that this modeling framework may
be applied more generally with alternative assumed error processes
dependent upon the requirements of a given application. For example, we
consider a normal error process in the simulation study that we present
in Section \ref{section6}.

\begin{figure}
  \centering
    \begin{tikzpicture}[x=1cm,y=1cm,
                        roundnode/.style={circle, draw=black, fill=white, thick, minimum size=.75cm},
                        lroundnode/.style={circle, draw=black, fill=black!10, thick, minimum size=.75cm},
                        squarednode/.style={rectangle, draw=black, fill=white, thick, minimum size=.75cm},
                        diamondnode/.style={diamond, draw=black, fill=white, thick, minimum size = .75cm}]
      % Nodes
      \node[diamondnode, scale = 1] (yij) {$\boldsymbol{y}_{ij}$} ; %
      \node[lroundnode, below = of yij, scale = 1] (lambdaij) {$\lambda_{ij}$} ; %
      \node[roundnode, left = of yij, scale = 1] (ti) {$\boldsymbol{t}_{i}$} ; %    \
      \node[squarednode, left = of ti, scale = 1] (sigma2t) {$\sigma^2_{t}$} ; %    \
      \node[roundnode, below = of ti, scale = 1] (thetai) {$\boldsymbol{\theta}_{i}$} ; %    \
      \node[squarednode, left = of thetai, scale = 1] (kappa) {$\kappa$} ; %
      \node[squarednode, below = of kappa, scale = 1, yshift = .5cm] (nu) {$\nu$} ; %
      \node[roundnode, above = of yij, scale = 1] (sigma2) {$\sigma^2$} ; %
      \node[squarednode, left = of sigma2, scale = 1] (D) {$D$} ; %
      \node[squarednode, above = of sigma2, scale = 1] (dsigma) {$d_{\sigma}$} ; %
      \node[squarednode, left = of dsigma, scale = 1, xshift = .5cm] (csigma) {$c_{\sigma}$} ; %
      \node[squarednode, below = of lambdaij, scale = 1] (N) {$N$} ; %
      \node[diamondnode, right = of yij, scale = 1, xshift = .5cm] (Xc) {$\boldsymbol{X}_{c}$} ; %
      \node[lroundnode, below = of Xc, scale = 1, yshift = -.05cm] (sj) {$\boldsymbol{s}_{j'}$} ; %
      \node[squarednode, below = of sj, scale = 1] (Dstar) {$D^*$} ; %
      \node[roundnode, right = of Xc, scale = 1] (gc) {$g_{c}$} ; %
      \node[roundnode, above = of gc, scale = 1] (gamma) {$\gamma$} ; %
      \node[roundnode, left = of gamma, scale = 1] (beta) {$\boldsymbol{\beta}$} ; %
      \node[roundnode, right = of gamma, scale = 1] (alpha) {$\alpha$} ; %
      \node[roundnode, below = of gc, scale = 1] (tau2) {$\tau$} ; %
      \node[squarednode, below = of tau2, scale = 1] (dtau) {$a_{\tau}$} ; %
      \node[squarednode, right = of dtau, scale = 1, xshift = -.5cm] (ctau) {$b_{\tau}$} ; %
      \node[squarednode, above  = of beta, scale = 1] (sigmabeta) {$\boldsymbol{\Sigma}_{\boldsymbol{\beta}}$} ; %
      \node[squarednode, left = of sigmabeta, scale = 1, xshift = .5cm] (mubeta) {$\boldsymbol{\mu}_{\boldsymbol{\beta}}$} ; %
      \node[squarednode, above = of gamma, scale = 1, xshift = -.5cm] (agamma) {$a_{\gamma}$} ; %
      \node[squarednode, above = of gamma, scale = 1, xshift = .5cm] (bgamma) {$b_{\gamma}$} ; %
      \node[squarednode, above = of alpha, scale = 1] (aalpha) {$a_{\alpha}$} ; %
      \node[squarednode, right = of aalpha, scale = 1, xshift = -.5cm] (balpha) {$b_{\alpha}$} ; %
      \node[squarednode, right = of alpha, scale = 1, yshift = .5cm] (calpha) {$c_{\alpha}$} ; %
      \node[squarednode, right = of alpha, scale = 1, yshift = -.5cm] (dalpha) {$d_{\alpha}$} ; %
      \node[roundnode, right = of gc, scale = 1] (delta) {$\delta$} ; %
      \node[roundnode, below = of delta, scale = 1] (omega) {$\omega$} ; %
      \node[ below = of thetai , xshift = -.55cm, yshift = .65cm]  (i) { $i$ };
      \node[ below = of lambdaij , xshift = -.6cm, yshift = 1cm]  (j) { $j$ };
      \node[ below = of sj , xshift = -.5cm, yshift = 1.05cm]  (jprime) { $j'$ };
      \node[ below = of Xc , xshift = -.65cm, yshift = .85cm]  (c) { $c$ };
      \node[squarednode, right = of delta, scale = 1] (sigma2_delta) {$\sigma^2_{\delta}$} ; %
      \node[squarednode, right = of omega, scale = 1] (b_omega) {$b_{\omega}$} ; %

      \edge {kappa} {thetai} ;
      \edge {nu} {thetai} ;
      \edge {sigma2t} {ti} ;
      \edge {ti} {yij} ;
      \edge {lambdaij} {yij} ;
      \edge {sigma2} {yij} ;
      \edge {D} {yij} ;
      \edge {thetai} {yij} ;
      \edge {sj} {yij} ;
      \edge {csigma} {sigma2} ;
      \edge {dsigma} {sigma2} ;
      \edge {N} {sj} ;
      \edge {Dstar} {sj} ;
      \edge {N} {lambdaij} ;
      \draw[dashed,->] (sj) -- (Xc);
      \edge {Xc} {gc} ;
      \draw[dashed,->] (lambdaij) -- (gc);
      \edge {tau2} {gc} ;
      \edge {alpha} {gc} ;
      \edge {gamma} {gc} ;
      \edge {beta} {gc} ;
      \edge {sigmabeta} {beta} ;
      \edge {mubeta} {beta} ;
      \edge {bgamma} {gamma} ;
      \edge {agamma} {gamma} ;
      \edge {aalpha} {alpha} ;
      \edge {balpha} {alpha} ;
      \edge {calpha} {alpha} ;
      \edge {dalpha} {alpha} ;
      \edge {dtau} {tau2} ;
      \edge {ctau} {tau2} ;
      \edge {delta} {gc} ;
      \edge {omega} {gc} ;
      \edge {b_omega} {omega} ;
      \edge {sigma2_delta} {delta} ;

       \plate[inner sep=.25cm, color=black] {j} { (yij) (lambdaij) } {  };
    \plate[inner sep=.3cm, color=black] {i} { (yij) (lambdaij) (ti) (thetai) } {  };
    \plate[inner sep=.25cm, color=black] {jprime} { (sj) } {  };
    \plate[inner sep=.25cm, color=black] {C} { (Xc)  (gc) } {  };

\end{tikzpicture}
\caption{Plate diagram for the two-stage record linkage and downstream growth model. Where $i=1,2$ denotes the file index, $j = 1,\dots,n_i$ denotes the record index within file $i$, $j' = 1,\dots,N$ denotes the latent location index, and $c = 1, \dots, \lvert \mathcal{C}^G\left(\boldsymbol{\Lambda}\right) \rvert$ denotes the growth cluster index. The round nodes indicate parameters, while square nodes indicate hyperparameters. We note that solid arrows denote stochastic relationships, while dashed arrows identify the inputs from the record linkage model to the downstream growth model. This framework provides the structure for the LA approach discussed in detail in Section \ref{section4}.}
\label{plate_diag}
\end{figure}
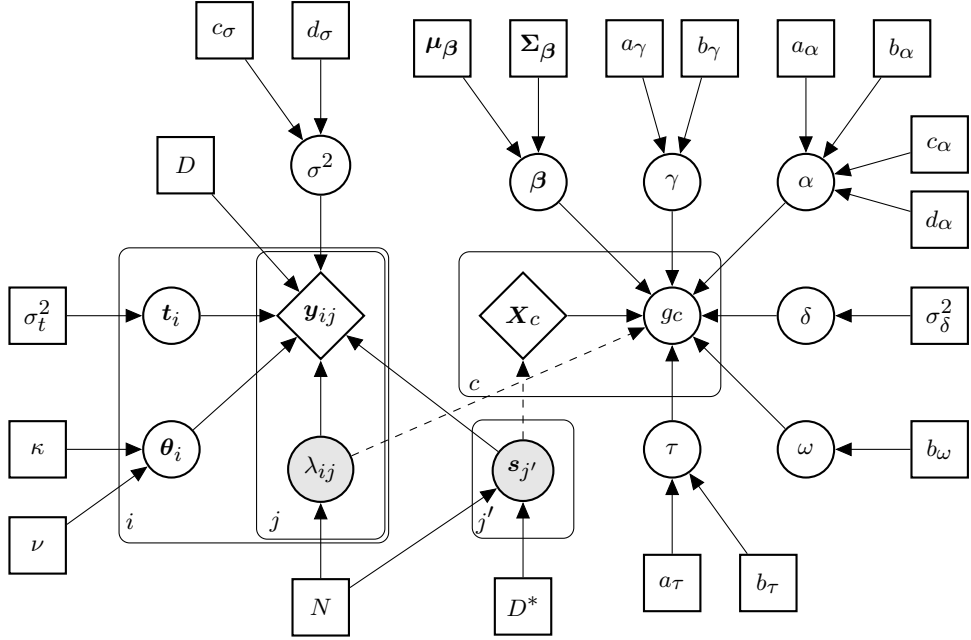

Combining the spatial record linkage and downstream growth models, as
seen in the plate diagram in Figure \ref{plate_diag}, we obtain the
structure for the two-stage modeling approach. We would like to
emphasize the distinction between the two models and the assumptions
that are made in each. The linkage model is designed to be as flexible
as possible to identify the relationship between the observed records
and the latent locations, while the growth model is designed to estimate
the growth of the trees. The models are specified to be used together,
but the assumptions made in the linkage model are not necessarily
identical to those made in the growth model. This two-stage approach
allows the linkage to be used for a variety of downstream modeling
objectives without the need to rerun the linkage model for each task.
Following the LA procedure outlined by
\cite{sadinleBayesianPropagationRecord2018a}, we propose using a random
sample of iterations from the marginal posterior of the linkage
structure \(\bmLambda\) and the latent spatial point process \(\bms\),
obtained from the linkage stage, as inputs for the downstream model. The
LA approach effectively marginalizes out the uncertainty from the
linkage and the latent locations and provides equivalent inference for
the growth model parameters compared to the marginal inference that
would be obtained from a joint model under certain conditions. We
discuss the requirements and justification for this approach in greater
depth in Section \ref{section4}.

\subsection{Computational Strategies}\label{section3.3}

The two-stage Bayesian hierarchical model framework that we propose has
many strengths including interpretability, the ability to incorporate
relevant prior knowledge, and robust uncertainty quantification across
the entire modeling pipeline. However, Bayesian record linkage modeling
approaches are known to carry additional computational overhead that can
be prohibitive to the use of these models in practice when dealing with
large datasets (see \cite{steortsComparisonBlockingMethods2014} and
\cite{marchantDblinkDistributedEndtoEnd2021} for more complete
discussions). We alleviate some of the computational expense associated
with the use of a Markov chain Monte Carlo Gibbs sampling algorithm for
our model through a few key mechanisms discussed below.

\begin{figure}
\centering
\includegraphics[scale = .5]{"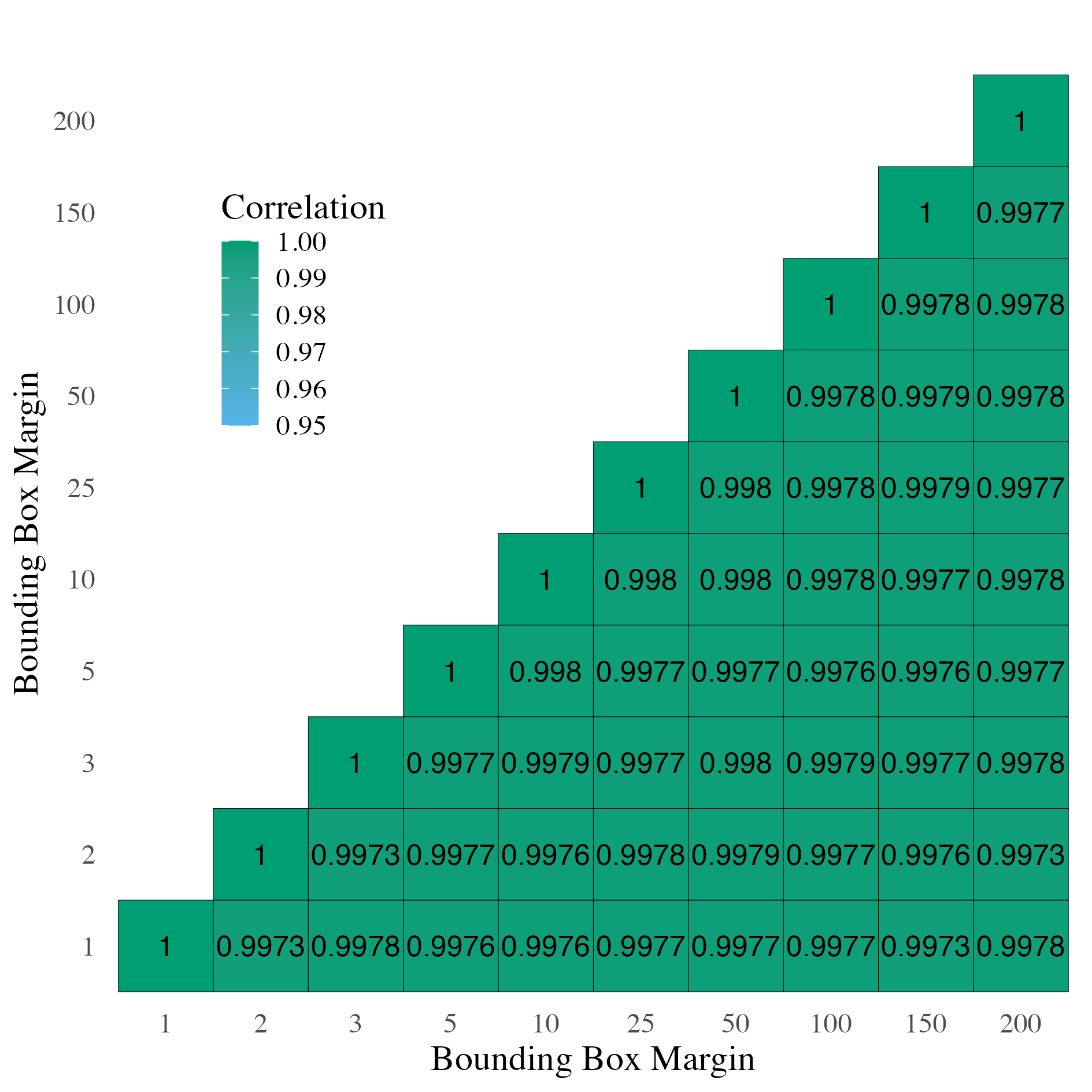"}
\caption{Correlation heatmap for the posterior similarity cluster scores for varying bounding box margins for an area of size \qty[mode = text]{200}{\unit{\meter^2}}. The bounding box margin value specifies the distance to the boundary from the observed point. For example, a margin of \qty[mode = text]{2}{\unit{\meter}} corresponds to a \qty[mode = text]{16}{\unit{\meter^2}} bounding box centered at the the observed location of the individual.}
\label{psm_cor}
\end{figure}

One of the most common approaches for improving the scalability of
record linkage models is to exact some form of deterministic blocking
for records to reduce the number of comparisons necessary. This
mechanism is commonly employed across both Bayesian and frequentist
record linkage implementations as a preprocessing step that invalidates
certain linkages that are deemed to be implausible
(\cite{steortsComparisonBlockingMethods2014, murrayProbabilisticRecordLinkage2015}).
While certain deterministic blocking schemes may impact the accuracy of
the linkage and fail to adequately quantify the uncertainty associated
with the procedure, we are able to take advantage of the spatial
structure of our data and the biological limitations that invalidate
certain links between observed records and \(\bms_{j'}\) as a function
of euclidean distance. As an alternative to blocking, we implement a
sampling scheme for \(\bmLambda\) in our Gibbs sampling algorithm that
allows us to approximate the posterior linkage structure under the
assumption that the observed location for an individual must be within a
maximum distance of the true latent location \(\bms_{j'}\) that it is
associated with. In contrast to blocking schemes which invalidate links
as a function of comparisons between records, our approach limits the
linkage structure directly. Absent the use of a blocking or
approximation scheme, the time required to sample from the true
posterior distribution of the linkage structure \(\bmLambda\) increases
quadratically with the number of records. Instead of considering the
full set \(\bms\) of possible latent locations for each record when
sampling the latent matching structure, we consider only \(\bms_{j'}\)
within a bounding box around the observed record. Additionally, we
impose the restriction that there must be at least two candidate
\(\bms_{j'}\) within the bounding box otherwise we increase the size of
the box iteratively until this condition is met to ensure a reasonable
approximation of the cluster assignment probabilities. We note that the
spatial bounding approach yields samples from an approximate posterior
distribution, however the \(\bms_{j'}\) removed from consideration have
near zero probability associated with them as possible matches and their
removal allows us to maintain a consistent computational cost in the
sampling of each individual \(\lambda_{ij}\). In Figure \ref{psm_cor},
we consider the correlation between posterior similarity scores, which
measure how often records are estimated to be coreferent, for bounding
boxes of varying sizes. We see that the correlations between posterior
similarity scores are close to 1 across bounding box sizes,
demonstrating the accuracy of the spatial bounding box approach for
approximating the true posterior distribution of the linkage structure
for moderately dense subsets of the empirical data.

\begin{figure}
\centering
\includegraphics[width = 400pt]{"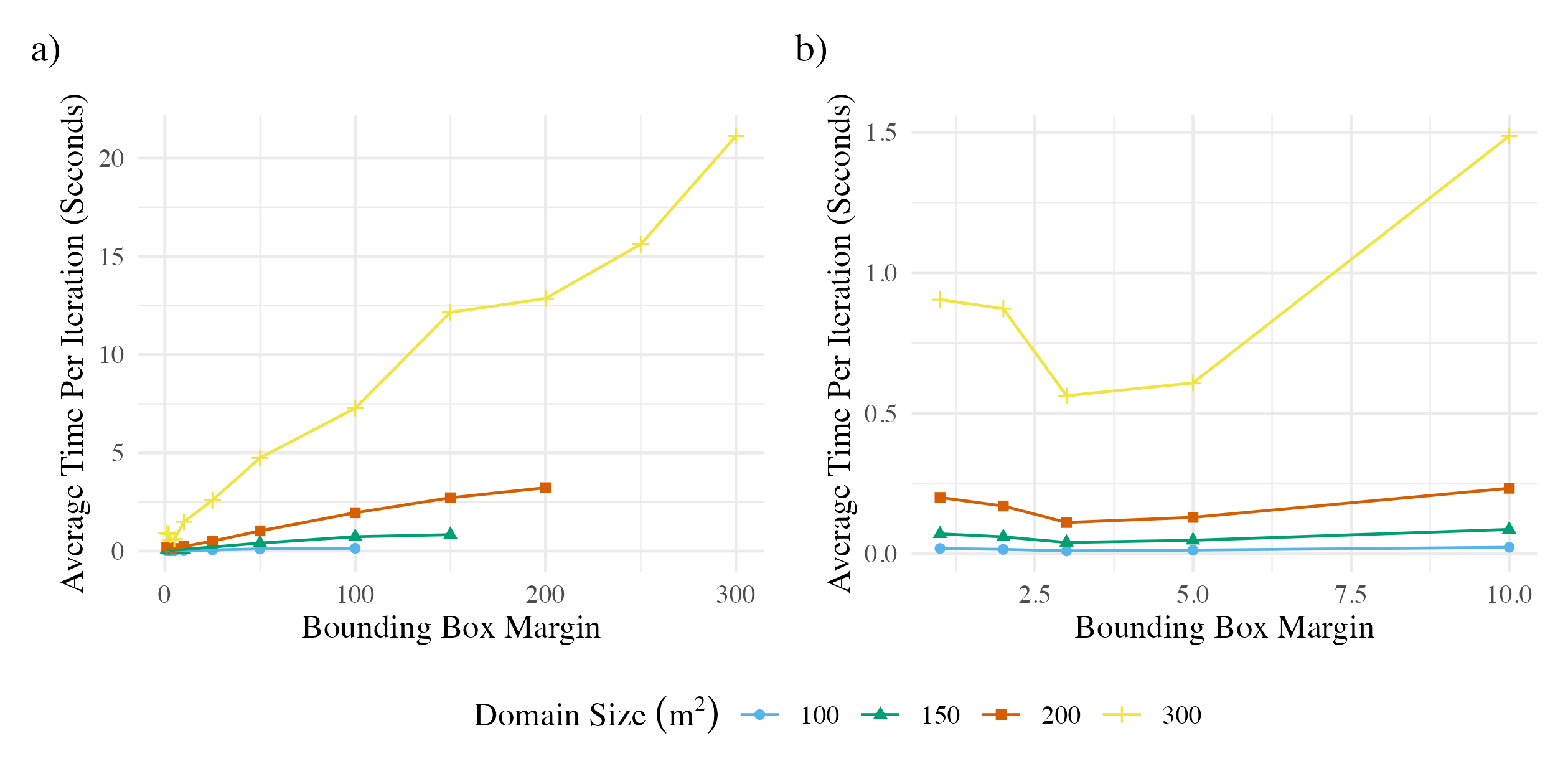"}
\caption{Record linkage model Gibbs sampler timing results per iteration for varying bounding box margins around each sampled point. We consider the timing for areas of size \qty[mode = text]{100}{\unit{\meter^2}}, \qty[mode = text]{150}{\unit{\meter^2}}, \qty[mode = text]{200}{\unit{\meter^2}}, and \qty[mode = text]{300}{\unit{\meter^2}}. Plot a) shows the full timing results, while plot b) shows an inset for smaller bounding box margins.}
\label{timing_plot}
\end{figure}

Functionally, this scheme improves both the speed and efficiency of the
record linkage model, as seen in Figure \ref{timing_plot}, allowing the
model to scale to much larger domains of interest, which is a clear
limitation of alternative modeling approaches. As the size of the
bounding box increases, we observe a near exponential increase in the
average time required per iteration of the sampler run on a dense
\qty[mode = text]{300}{\unit{\meter^2}} subset of our empirical dataset.
Over this domain, we observe a 97.3 times speedup per iteration on
average when using a bounding box of \qty[mode = text]{3}{\unit{\meter}}
with the resulting linkage having a posterior similarity score
correlation of .9984 compared to not using a bounding box. However, we
note that the speed improvement for decreasing bounding box sizes is not
universal as at some point we are required to expand the size of the box
iteratively to meet the conditions established for guaranteeing a
reasonable approximation of the cluster assignment probabilities.

To optimize the raw computation speed, the MCMC sampler is written in R
and C++ using \texttt{Rcpp}
(\cite{eddelbuettelRcppSeamlessIntegration2023}) and
\texttt{RcppArmadillo}
(\cite{eddelbuettelRcppArmadilloRcppIntegration2023}) to improve
scalability over a base R implementation. We implement the downstream
growth model in \texttt{rstan} to take advantage of the speed and
flexibility of the NUTS algorithm
(\cite{standevelopmentteamRStanInterfaceStan2023}). Additionally, the
optimized parallel computation available in \texttt{rstan} reduces the
time required to fit the downstream growth model with minimal additional
architecture required. The details of our Gibbs sampling algorithm for
the spatial record linkage model may be found in the Supplementary
Materials in Appendix A (\cite{drewBayesianRecordLinkage2025sup}).

\section{Linkage-Averaging for Parameters from Auxiliary Data Models}\label{section4}

We present a discussion of the theoretical justification for the LA
approach, introduced by \cite{sadinleBayesianPropagationRecord2018a} for
population size estimation, for a general downstream task with auxiliary
data, i.e.~regression, when paired with a record linkage model that
models the observed records as noisy versions of a set of true latent
field values as in \cite{steortsBayesianApproachGraphical2016} and
\cite{liseoBayesianEstimationPopulation2011a} and as defined in Section
\ref{section3.1}. We demonstrate that under the following two mild
conditions, this LA approach may be reframed to provide proper Bayesian
inference for the parameters of a more general downstream task.

\begin{condition}\label{c1}
Our beliefs regarding the linkage structure $\bmLambda$ and the true latent field values $\bms$ are quantified by the joint posterior distribution $p_\text{LRL} \left(\bmLambda, \bms \mid \bmy \right)$, arising from a record linkage model employing a latent matching structure, where the posterior is proportional to the product of the likelihood $\mathcal{L}_\text{LRL}\left(\bmLambda, \bms \mid \bmy \right)$ and the joint prior $p\left(\bmLambda, \bms\right)$.
\end{condition}

We note that this condition depends on the use of a record linkage model
that employs a latent matching structure in which the fields are modeled
directly, though it would be straight forward to adapt for an
alternative linkage model construction.

\begin{condition}\label{c2}
If the true linkage structure $\bmLambda$ and latent field values $\bms$ were known, the posterior $p_\text{AD} \left(\bmTheta \mid \mathcal{C}\left(\bmLambda\right), \bmX\left(\bms\right)\right)$ arising from an auxiliary data model with likelihood $\mathcal{L}_\text{AD}\left(\bmTheta \mid \mathcal{C}\left(\bmLambda\right), \bmX\left(\bms\right)\right)$ and joint prior $p\left(\bmTheta\right)$, which may be further decomposed depending on the structure of the model, would encapsulate our beliefs regarding the downstream model parameters.
\end{condition}

The second condition describes the inferential process for \(\bmTheta\),
the vector of parameters from the auxiliary data model, under the
assumption that the true linkage structure and latent field values are
known. The combination of these two conditions provide the basis for our
underlying argument, such that if these conditions hold, the following
relationship \begin{equation}
  p_{\text{LA}} \left(\bmTheta\right) = \E_{\bmLambda,\bms \mid \bmy} \left[ p_\text{AD} \left(\bmTheta \mid \mathcal{C}\left(\bmLambda\right), \bmX\left(\bms\right)\right) \right] = \sum_{\bmLambda} \sum_{\bms} p_\text{AD} \left(\bmTheta \mid \mathcal{C}\left(\bmLambda\right), \bmX\left(\bms\right)\right) \mathcal{L}_\text{LRL}\left(\bmLambda, \bms \mid \bmy\right), \nonumber
\end{equation} is obvious to consider given its clear interpretation. We
also demonstrate that \(p_{\text{LA}} \left(\bmTheta\right)\) is a
proper posterior distribution. In order to perform inference on
\(\bmTheta\) and \(\left(\bmLambda, \bms\right)\) given \(\bmy\), we
require a joint prior for \(\left(\bmTheta, \bmLambda, \bms\right)\)
such that \begin{equation}
  p\left(\bmTheta, \bmLambda, \bms\right) = p_\text{AD} \left(\bmTheta \mid \mathcal{C}\left(\bmLambda\right), \bmX\left(\bms\right)\right)  p\left(\bmLambda\right) p\left(\bms\right), \nonumber
\end{equation} which follows naturally from Conditions \ref{c1} and
\ref{c2} above.

\begin{theorem}[Bayesian validity of linkage-averaged auxiliary data model parameters joint posterior]\label{thm1}
  The marginal posterior of $\bmTheta$ under the likelihood $\mathcal{L}_\text{LRL}\left(\bmLambda, \bms \mid \bmy\right)$ of the latent record linkage model and joint prior $p_\text{AD} \left(\bmTheta \mid \mathcal{C}\left(\bmLambda\right), \bmX\left(s\right)\right) p\left(\bmLambda\right) p\left(\bms\right)$ is $p_{\text{LA}}\left(\bmTheta\right)$.
\end{theorem}

Theorem \ref{thm1} establishes \(p_{\text{LA}}\left(\bmTheta\right)\) as
a valid posterior distribution. We provide the proof for \ref{thm1} in
Appendix B of the Supplementary Materials
(\cite{drewBayesianRecordLinkage2025sup}), as the details are similar to
the proof of \cite{sadinleBayesianPropagationRecord2018a}. We note that
the proof of \cite{sadinleBayesianPropagationRecord2018a} holds
specifically for population size estimation, in comparison to the result
for a general downstream task with auxiliary data which we have
established. In practice, we approximate the linkage-averaged posterior
of \(\bmTheta\), \(p_{\text{LA}}\left(\bmTheta\right)\), with a random
sample
\(\bmTheta^{\left(1,t\right)}, \dots, \bmTheta^{\left(l,t\right)} \sim p_\text{AD} \left(\bmTheta \mid \mathcal{C}\left(\bmLambda\right)^{\left(t\right)}, \bmX\left(\bms\right)^{\left(t\right)}\right)\),
for each \(t = 1,\dots,k\), such that \begin{equation}
p_{\text{LA}}\left(\bmTheta\right) \approx \frac{1}{kl} \sum_{t = 1}^{k} \sum_{u = 1}^{l} I\left(\bmTheta = \bmTheta^{\left(u, t\right)}\right), \nonumber
\end{equation} where \(k\) and \(l\) are chosen to be sufficiently large
to provide a reasonable approximation to the true posterior.

The spatial record linkage and downstream growth models detailed in
Sections \ref{section3.1} and \ref{section3.2} respectively, clearly
satisfy the construction discussed above with
\(\bmTheta = \left(\alpha, \gamma, \bmbeta, \tau, \delta, \omega \right)\)
and where \(\bmX\left(\bms\right)\) represents the auxiliary data
component of the model. We note that an alternative to the LA approach
is to model the record linkage and downstream task jointly, which allows
the downstream task to inform the file linkage procedure. For example,
\cite{gutmanBayesianProcedureFile2013} discuss a joint modeling approach
based on multiple imputation that iteratively samples the unknown
linking partition and the downstream model parameters. In their
framework, the unknown links are treated as missing data and imputed.
While the joint modeling approach may potentially improve the linkage,
it is often accompanied by a substantially increased computational
burden and the performance is sensitive to model misspecification for
the downstream model. In contrast, the LA framework that we present
provides equivalent marginal inference for the downstream model
parameters as that obtained from a joint model, under the assumptions of
Conditions 1 and 2, and allows more flexibility for the researcher to
recycle the linkage for multiple downstream tasks of interest that may
be implemented in parallel in a straight forward and efficient fashion.

\section{Estimation of Annual Growth Curves for Rocky Mountain Conifer Forests}\label{section5}

In this section, we return to the empirical data and related hypotheses
regarding the annual growth behavior of Southern Rocky Mountain conifer
forests presented in Section \ref{section2}. We employ the two-stage LA
approach for estimating the downstream growth model parameters using
\(k = 100\) randomly sampled iterations from the joint posterior
distribution of the linkage structure \(\bmLambda\) and latent locations
\(\bms\) as the input for the growth model. For each pair
\(\left( \bmLambda^{(k)}, \bms^{(k)} \right)\), we derive the set of
growth clusters, \(\mathcal{C}^G\left(\bmLambda^{(k)}\right)\), and the
set of location dependent covariates,
\(\bmX \left(\bms_{\mathcal{C}^G\left(\bmLambda^{(k)}\right)}^{(k)} \right)\),
which are then used to fit the growth model defined in Section
\ref{section3.2}. This procedure allows us to obtain estimates of the
marginal posterior distributions of the growth model parameters of
interest that are equivalent to the marginals obtained from a joint
model for the linkage structure, \(\bms_{j'}\), and the growth model
parameters, as discussed in Section \ref{section4}. For this analysis,
we specify \(r_1 = .9\) and \(r_2 = 1.6\) such that we consider
primarily positive growth with a maximum increase of 60\% of the initial
observed canopy volume over the 4 year study period. These cutoffs
reflect typical growth behavior and disqualify implausible clusters
arising from errors in the linkage and due to environmental mechanisms
like damage from extreme wind or lightning that do not reflect the
biological mechanisms of tree growth.

In addition to the topographic covariates discussed in Section
\ref{section2}, we also include three inter-tree competition metrics.
\cite{fagerbergIndividualtreeDistancedependentGrowth2022} highlight the
importance of including competition indices in individual tree growth
models for conifer species. For our application, we consider relative
spacing index (RSI), the ratio of the nearest neighbor distance to the
average neighbor distance, larger neighbor volume (LNV), the summed
canopy volumes of a tree's larger neighbors, and neighborhood density
(ND), the density of individuals within the neighborhood of the
individual. All competition metrics are calculated using the observed
locations from the first scan (2015) within a
\qty[mode = text]{15}{\unit{\meter}} neighborhood around each point such
that all three are considered semi-distance dependent competition
indices. \cite{maQuantifyingIndividualTree2018} calculate LiDAR derived
tree competition indices using a \qty[mode = text]{15}{\unit{\meter}}
neighborhood, and we adopt the same neighborhood size for our analysis.
To ensure that these metrics are accurate for all points considered, the
downstream growth model is only fit to growth clusters located more than
\qty[mode = text]{15}{\unit{\meter}} from the boundary of the study
domain. We note that RSI and ND are measures of symmetric competition
while LNV captures asymmetric competition among individuals to account
for the variation possible across the range of competitive effects. An
in depth discussion of competition indices and their construction may be
found in \cite{pommereningTammReviewTree2018} and
\cite{contrerasEvaluatingTreeCompetition2011}.

Given the size of the study domain
(\textasciitilde{}\qty[mode = text]{2}{\unit{\km\squared}}), we fix the
rotation parameter for the second scan, \(\theta_2\), to be zero as even
a very small degree of rotation can have a large effect on points near
the boundary of the domain. We allow for the possibility of scan-wide
translation in this analysis, and note that the choices of which image
alignment components to include and the strength of their constituent
priors will likely depend on the application. We select noninformative
and weak priors, where appropriate, for the record linkage and
downstream growth model parameters according to the outline in Section
\ref{section3}. We specify \(q=1.25\) in determining the maximum number
of unique latent individuals \(N\) to provide flexibility across the
range of point densities observed in the study area. The convergence of
the record linkage model and downstream model variants were assessed by
examining traceplots and Gelman Rubin statistics
(\cite{gelmanInferenceIterativeSimulation1992a}) for each approach. The
full model specification details and select convergence diagnostics may
be found in Appendix C of the Supplementary Materials
(\cite{drewBayesianRecordLinkage2025sup}).

\begin{figure}
\centering
\includegraphics[width = 400pt]{"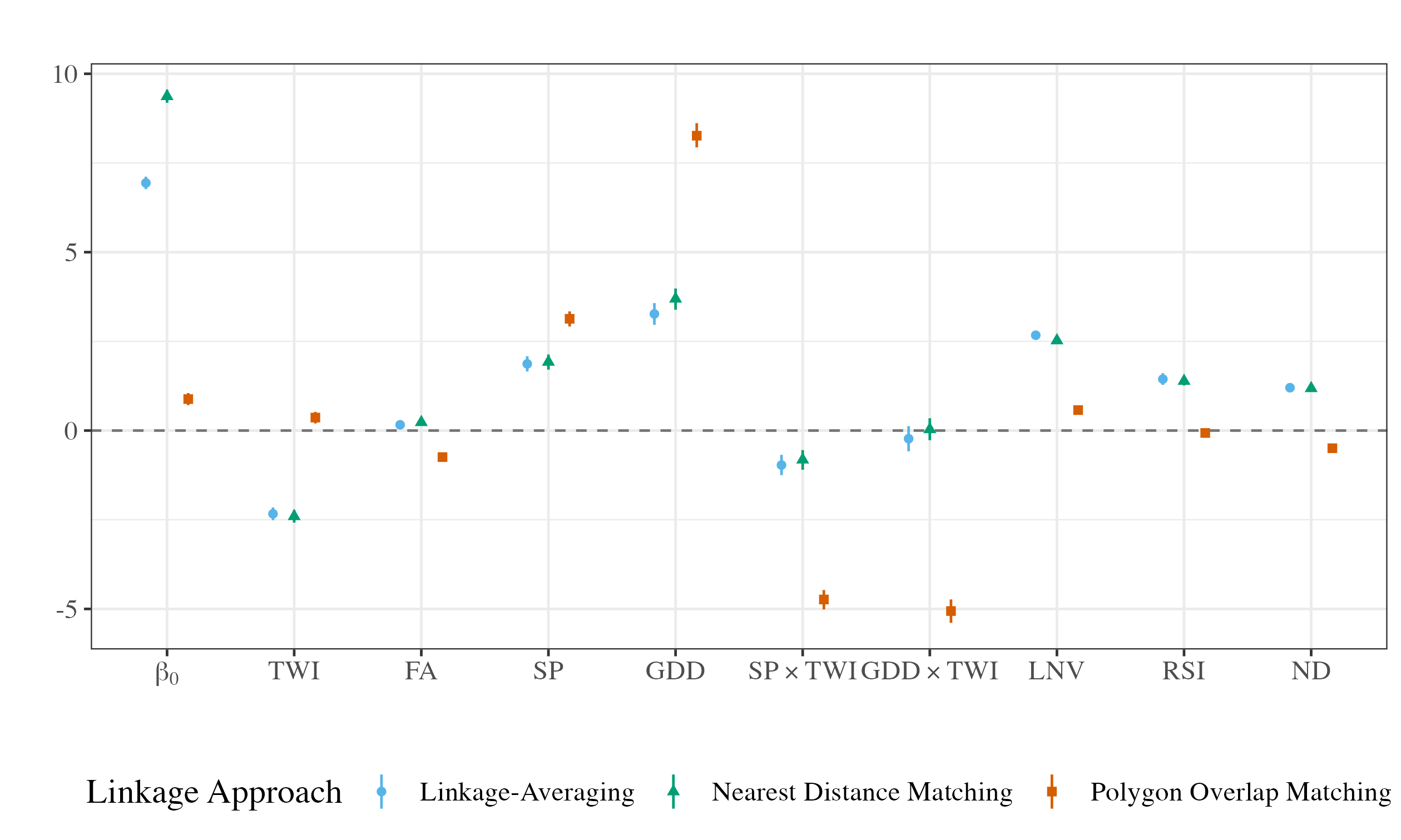"}
\caption{Comparison of the 90\% credible intervals for the growth asymptote and the topographic and competition metric covariate coefficients obtained from the downstream model fit using data derived from the three different linkage approaches (LA, NDM, and POM).}
\label{coverage_plot}
\end{figure}

In concert with the two-stage LA approach, we consider two alternative
heuristic strategies for linking trees across scans which we term
nearest distance matching (NDM) and polygon overlap matching (POM). The
NDM algorithm matches each tree crown location from the first scan with
the closest point from the second scan. The POM approach uses the
derived crown geometries from the LiDAR scans and traces the crown
polygons from the 2015 scan forward and considers the overlap with the
polygons from the 2019 scan and the change in canopy volume is
calculated as the difference between the estimated volumes. While these
methods do not perform deduplication and fail to provide uncertainty
quantification for the linkage, they represent simple and easy to
implement strategies for identifying unique individuals across scans and
obtaining growth estimates that are analogous to methods being used in
practice. For example, \cite{maQuantifyingIndividualTree2018} use a more
sophisticated heuristic matching algorithm coupled with manual review of
marginal matches. We apply the same growth cluster restrictions for
these methods as for the LA approach, so the set of derived growths is
characteristically equivalent across the three linkage procedures.
During model fitting, we considered 4 candidate growth models, 3 with
the non-linear Michaelis-Menten mean function, for each linkage strategy
(Skewed t, Skew Normal, Normal, and Multiple Linear Regression with
Normal Errors) and evaluated the model fit using the Continuous Ranked
Probability Score (CRPS) as suggested by
\cite{delacruzBayesianAnalysisNonlinear2009} following the discussion of
\cite{gneitingStrictlyProperScoring2007}. We used the scaled CRPS
(sCRPS) presented by \cite{bolinLocalScaleInvariance2023} which has been
shown to be locally scale invariant and an improvement over the standard
CRPS for model selection. The Skewed t model discussed in Section
\ref{section3} was identified by the sCRPS metric as the top performing
model for all 3 of the linkage schemes (additional details may be found
in Appendix C of the Supplementary Materials
(\cite{drewBayesianRecordLinkage2025sup})). Figure \ref{coverage_plot}
displays the 90\% credible intervals for the covariate coefficients
obtained from the three linkage approaches for the Skewed t model.

\begin{figure}
\centering
\includegraphics[width = 400pt]{"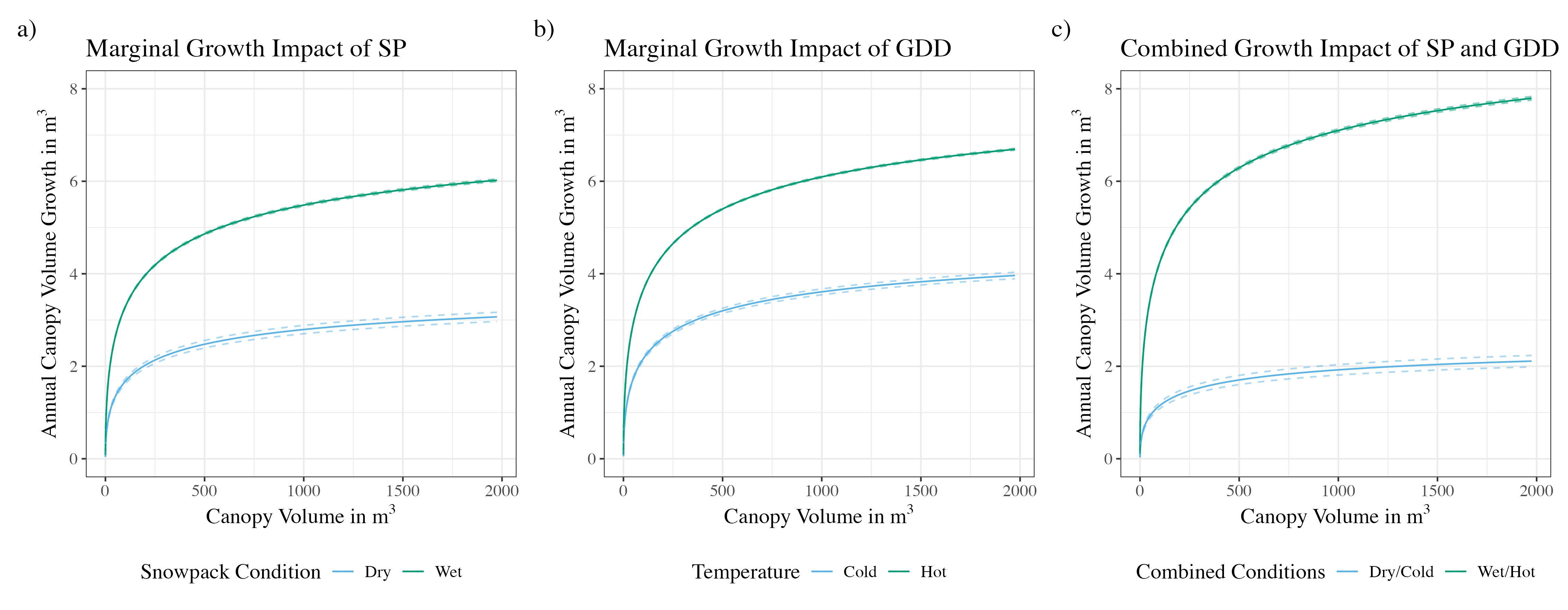"}
\caption{Comparison of the estimated growth curves for varying quantiles of covariates of interest where plot a) is the growth curve for Snowpack Persistance, plot b) is the growth curve for Growing Degree Days, and plot c) is the growth curve for Snowpack Persistance and Growing Degree Days simultaneously. The plots demonstrate the change in growth behavior for low and high quantiles of the covariates while holding all other covariates at their median values, where the designations Dry/Cold and Wet/Hot correspond to the 20th and 80th quantiles respectively across the plots.}
\label{growth_curve_plot}
\end{figure}

We see from the coverage plot that the POM approach results in
drastically different estimates for the coefficients in terms of
magnitude, and in some instances sign, coupled with high degrees of
certainty in the estimates. In contrast, the NDM and LA approaches
produce more similar estimates for the coefficients given that they are
both distance based linkage approaches, although the NDM is
deterministic and so does not marginalize out the uncertainty from the
linkage procedure when fitting the downstream task like the LA approach.
We also note that the estimates for the growth asymptote \(\beta_0\) are
notably different across the three linkage approaches, even for models
with similar estimates of the covariate coefficients. In Figure
\ref{growth_curve_plot}, we provide a comparison of the estimated size
dependent growth curves arising from the LA model under high and low
growth scenarios as a function of the topographic covariates Snowpack
Persistence and Growing Degree Days, which are measures of water and
energy availability, respectively, as discussed in Section
\ref{section2}. The annual growth curve, \(\mu_c\) (Equation \ref{eq1}),
is a function of size, where the growth asymptote is adjusted by the
covariate values at the location of the tree. We consider the 20th and
80th quantiles of the empirical distribution for these covariates while
holding all other covariates at their median values to highlight the
marginal impact on growth for these individual covariates with 90\%
credible bands. In panel c), we examine the growth curves for both
covariates simultaneously to demonstrate the combined impact of the
covariates on growth behavior in both sub-optimal and optimal growth
conditions.

Our analysis suggests the importance of including environmental
variables related to growth conditions in addition to competition
indices in modeling size-dependent individual tree growth over large
spatial domains
(\cite{fordCompetitionAltersTree2017, maesEnvironmentalDriversInteractively2019}).
While the growth behavior is primarily constrained by size in our
analysis, these additional metrics are influential in determining the
growth behavior of forests across varied terrains and localized
densities. Our work reinforces other studies on environmental
constraints to conifer growth in the region which emphasize that both
available energy and water from snowpack are important growth
constraints
(\cite{berkelhammerPersistencePlasticityConifer2020, carroll_efficiency_2020}).
Somewhat surprisingly, we observed negative effects of a soil moisture
proxy (HAS Wetness Index, Figure \ref{emp_rast_plot} panel d) on tree
growth, indicating that growth of some trees in our study domain may be
limited in the wettest soils (\cite{marksVariationTreeGrowth2020}). We
also observe that the interaction of topographic proxies for energy and
water availability may have an impact on growth when accounting for some
collection of symmetric and asymmetric competition metrics. This
inference for the downstream model is sensitive to the choice of linkage
approach, particularly when considering the estimated growth asymptote.

\section{Simulation Results}\label{section6}

In this section, we perform a sequence of simulation studies to examine
the efficacy of our modeling framework. In Section \ref{section6.1}, we
introduce the data generation algorithm for producing biologically
realistic simulated data sets. Subsequently, in Section
\ref{section6.2}, we assess the performance of the LA model as applied
to a collection of simulated datasets under various scenarios modeled
after the empirical data discussed in Section \ref{section2}.

\subsection{Data Simulation}\label{section6.1}

Historically, ecological surveys of forest growth dynamics have relied
extensively on field measurement data for validating models using ALS
data, as in \cite{maQuantifyingIndividualTree2018}. These field surveys
are often time consuming and expensive to perform, and provide a limited
characterization of the model performance across a wide range of
scenarios. Due to the scale and complexity of the study area we are
considering, a validation dataset is unavailable. Instead, we gauge the
efficacy of our model by considering the performance on simulated data
across a variety of possible conditions as motivated by our empirical
data. The majority of the existing simulation frameworks for marked
point processes assume independence between the spatial point process
and the mark distribution as noted by
\cite{yongtaoTestIndependenceMarks2007}, however our application
necessitates location dependent marks which are specified to be canopy
volumes.

We address the disconnect between the available off the shelf methods
and the requirements of our application through the use of a data
simulation algorithm constructed to approximate the underlying marked
spatial point process and the relevant biological mechanisms of forest
populations, such as growth and recruitment, using three subjectively
selected subsets of the empirical data with varying point densities as a
basis. We initialize the procedure by simulating the latent point
process \(\bms\) with a modification of the modeling scheme of
\cite{mollerMechanisticSpatiotemporalPoint2016a} for marked point
processes, in order to include topographically derived covariates in the
simulation of the mark distribution. We consider three point densities
motivated by the range of densities observed in the RMBL dataset. The
point densities are .04, .06, and .08 individuals per square meter,
which we describe as low, medium, and high, respectively, throughout the
remainder of this section. We note that the simulated data is
constructed to generate data from two files corresponding to the
empirical data in our application.

One of the key innovations of the
\cite{mollerMechanisticSpatiotemporalPoint2016a} approach is to equate a
marked spatial point process with a spatio-temporal point process by
ordering the marks and treating them as arrival times in a
spatio-temporal process. For each selected density, we use the observed
sizes from the 2015 dataset to generate the approximate arrival times of
the points and then predict the mark associated with each point as a
function of time, neighborhood characteristics, and topographic
covariates in an iterative fashion until we obtain a point realization
matching the intensity of the empirical reference pattern. We employ an
embedded gradient boosted tree model, built using the \texttt{xgboost}
package (\cite{chenXgboostExtremeGradient2023}) in R, to predict the
marks given the set of derived features. The empirical data demonstrate
a notable pattern of inhibition, or regularity, at the
\qty[mode = text]{100}{\unit{\meter^2}} scale, so we include provisions
for interpoint interaction as a function of size in the data generation
procedure to capture this behavior. The interaction function for point
patterns defined by regularity, such as a Strauss process, are often
specified with a hard core radius such that points in the process cannot
be within a certain radius of each other
(\cite{leiningerBayesianAnalysisSpatial2014a}). In our process we assume
a soft core interaction radius such that we allow points to violate the
hard core interaction radius with low probability. All of the simulated
data is generated over \qty[mode = text]{130}{\unit{\meter^2}} areas,
and then restricted to the center
\qty[mode = text]{100}{\unit{\meter^2}} area to account for possible
edge effects in the point patterns. We use the raster images of the
topographic covariates, provided by RMBL, to draw the location specific
covariate values in order to simulate data that approximates the real
data as closely as possible.

To accurately reflect the biological mechanism of juvenile recruitment
(i.e.~the seeding of offspring trees) over time, we generate the number
and locations of potential recruits according to the realized parent
point process. Each parent point is assigned a number of recruits based
on its size and the locations of recruits are modeled as arising from a
\(\text{t}\left(1\right)\) distribution centered at the parent point.
The marks for recruits are simulated from a heavily right-skewed scaled
Beta distribution bounded by the minimum size observed in the
distribution of the parent points in order to capture the fact that a
relatively small number of recruits survive long enough to be
identified. We note the LiDAR process used to collect the empirical data
has a height detection threshold of approximately
\qty[mode = text]{2}{\unit{\meter}}, and so empirical data for the size
distribution of smaller trees was unavailable. However, the mechanics of
recruitment are well studied, and we incorporated relevant domain
knowledge in the construction of our mechanism allowing larger
individuals to seed more potential recruits using a sampling mechanism
incorporating the individual's proportion of the total biomass
contribution as the sampling weight. \cite{johnsonCanopyTreeDensity2021}
provide a thorough discussion of the mechanisms involved in recruitment
processes of conifer species which we leverage in our data generating
process.

\begin{figure}
\centering
\includegraphics[width = 400pt]{"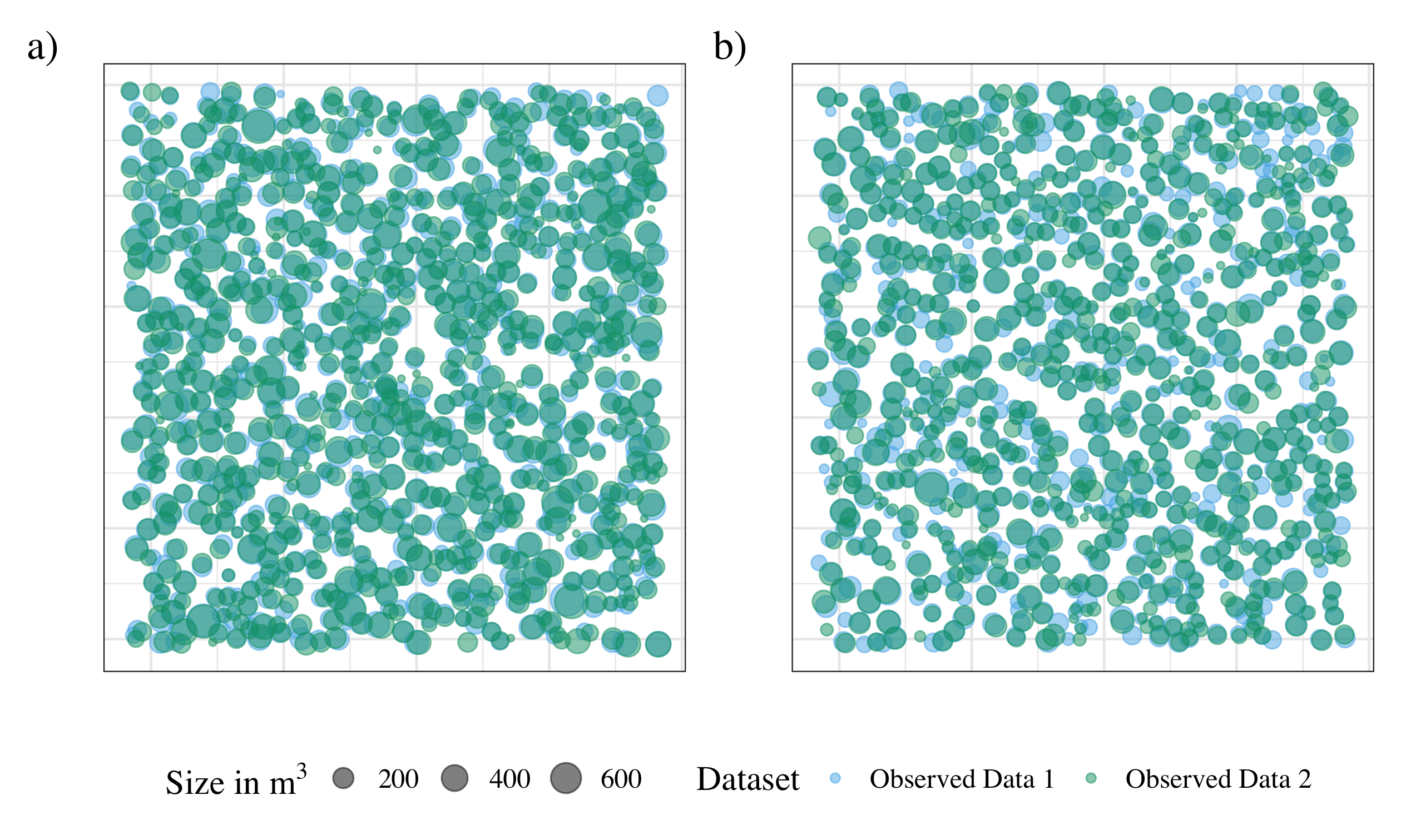"}
\caption{Comparison of the data from a medium density \qty[mode = text]{100}{\unit{\meter^2}} subset where plot a) is the simulated data and plot b) is the empirical data subset used to build the predictive model for the medium density validation data.}
\label{pp_comp_plot}
\end{figure}

We proceed in generating the observed data from two time points by
applying a simplified growth model to appropriate transformations of the
latent point configuration \(\bms\). For purposes of illustration, we
consider a growth model with the same mean structure as the model
presented in Section \ref{section3.2}. As noted before, our modeling
approach may be adapted for different error processes, and so for
simplicity we assume a standard Gaussian error process for this
simulation study instead of the Skewed t process discussed previously.
We introduce noise in the observed locations according to the data
process of the spatial record linkage model in Section \ref{section3.1}
such that each observed location is generated from a bivariate normal
distribution centered at the latent parent point \(\bms_{j'}\) with
measurement error \(\sigma^2\). Finally, we translate and rotate the
points to achieve the final spatial configurations for each file. Given
the observed marks for the points from the first file, for each point we
predict the growth from the first observed time to the second according
to the Michaelis-Menten style mean function \(\mu\) with measurement
error \(\tau^2\). We note that growth is also predicted for the
generated recruits, but without the measurement error component as these
points are technically unobserved in the first file due to their sizes
being below the detection threshold. The final set of observed data from
two files is obtained by truncating the generated patterns to the center
\qty[mode = text]{100}{\unit{\meter^2}} area. We apply the outlined
simulation framework to produce a collection of datasets with known
generating parameters and linkage structure as a baseline for assessing
model performance, in the absence of a field inventory validation
dataset, with varying levels of measurement error deemed to be
plausible. A comparison of the simulated and empirical data for a
\qty[mode = text]{100}{\unit{\meter^2}} medium density subset may be
seen in Figure \ref{pp_comp_plot}. A detailed discussion of the data
generation algorithm and underlying assumptions may be found in Appendix
D of the Supplementary Materials
(\cite{drewBayesianRecordLinkage2025sup}).

\subsection{Simulation Performance}\label{section6.2}

In this section, we assess the performance of the LA two-stage modeling
approach on data generated using the simulation scheme discussed in
Section \ref{section6.1}. We are able to gauge the efficacy of both the
record linkage model and the downstream growth model given that the true
linkage structure and parameter values used to generate the simulated
data are known. We consider the performance of the model on 100
simulated datasets from low, medium, and high densities, which
correspond to point intensities of approximately .04, .06, and .08
respectively over \qty[mode = text]{100}{\unit{\meter^2}} areas. We
first consider the performance of the spatial record linkage model and
then explore credible interval coverage rates for the growth model
parameters across 100 simulated data sets for each density and with
varying levels of noise in the observed locations.

\begin{figure}
\centering
\includegraphics[width = 400pt]{"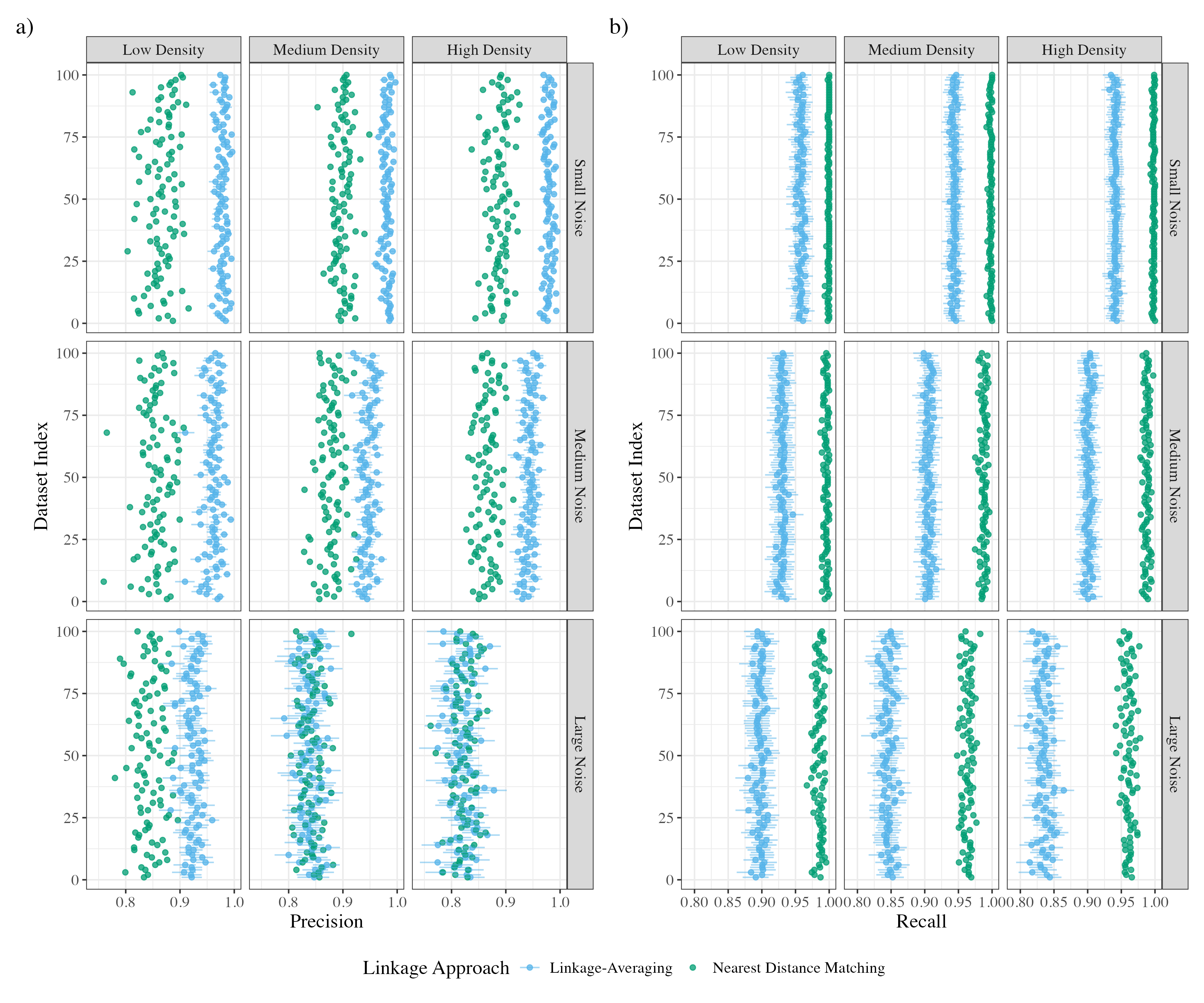"}
\caption{Plots comparing the precision a) and recall b) performance for the LA and NDM linkage approaches on 100 simulated datasets for each density and noise combination with known true linkage and with $\alpha = 1$.}
\label{prec_rec_plot}
\end{figure}

For the linkage model, we consider the metrics precision and recall,
which are standard evaluation criteria for classification tasks defined
as \begin{equation}
\text{Precision} = \frac{\text{TP}}{\text{TP} + \text{FP}} \quad \text{\&} \quad \text{Recall} = \frac{\text{TP}}{\text{TP} + \text{FN}}, \nonumber
\end{equation} where TP, FP, and FN are the number of true positives,
false positives, and false negatives respectively. Precision measures
the proportion of correctly identified matches out of all identified
matches while recall measures the proportion of correctly identified
matches out of all possible matches (\cite{alma991017200819703361}). We
summarize the linkage performance for \(\alpha = 1\) over 100 datasets
with known true linkage in Figure \ref{prec_rec_plot}. We considered
three noise levels for the generating process for the observed spatial
locations corresponding to \(\sigma = .25\), \(\sigma = .35\), and
\(\sigma = .45\) which we term small, medium, and large respectively.
The model was run with \(q=1.25\) when determining \(N\) to match the
value selected for the empirical data analysis. The results were similar
for two alternative \(\alpha\) values, which may be found in Appendix D
of the Supplementary Materials
(\cite{drewBayesianRecordLinkage2025sup}).

We note that the NDM algorithm only produces a single linkage estimate
for each dataset and consequently a single estimate of precision and
recall, while we obtain a full posterior distribution of the linkage for
each dataset from the spatial record linkage model. We see from Figure
\ref{prec_rec_plot} that the average precision performance for the
record linkage model tends to be a notable improvement over the NDM
approach across the varying levels of density and noise. We note that
the increase in precision for the record linkage model comes at a cost
in the form of reduced recall compared to the NDM approach. The NDM
approach matches each point from the first file with a point in the
second file, and resultingly captures more matches overall but the
accuracy of those matches is reduced relative to the LA approach. We
note that incorrect links function much like random noise and may result
in a form of attenuation bias, effectively driving the estimates of the
downstream model covariate coefficients to zero. By prioritizing
precision, we are able to draw more accurate conclusions about the
impacts of the covariates on growth for the trees that are linked.
However, this may result in additional bias when attempting to
generalize the findings beyond the observed data.

We next consider the performance of the downstream growth model in terms
of nominal coverage rates for 90\% credible intervals for the growth
model parameters of interest. We examine the same density, noise, and
\(\alpha\) settings as for the linkage model with growth parameter
values \(\gamma = 12\),
\(\bmbeta = \bbmx 3& .5& -.5& .5& -.5 \ebmx^\mT\), and \(\tau^2 = .5\).
Our prior specifications for overlapping parameters match those we use
in the real data application to provide as direct an analogue between
the two modeling scenarios as possible. The coverage results for the
growth model parameters over the 100 datasets for each setting may be
found in Table \ref{growth_cov_table}.

\begin{table}
\caption{Empirical coverage rates for 90\% credible intervals for covariate parameters of the downstream growth model across the true linkage, LA, and NDM linkage approaches. We consider the coverage over 100 datasets for each setting with $\alpha = 1$ where the bolded coverages are closest to the nominal level (excluding the true linkage).}
\label{growth_cov_table}
\centering
\begin{tabular}[t]{lllcccccccc}
\toprule
\multicolumn{3}{c}{\textbf{ }} & \multicolumn{8}{c}{\textbf{Empirical Coverage by Parameter}} \\
\cmidrule(l{3pt}r{3pt}){4-11}
Density & Noise & Linkage Approach & $\alpha$ & $\beta_0$ & $\beta_1$ & $\beta_2$ & $\beta_3$ & $\beta_4$ & $\gamma$ & $\tau^2$\\
\midrule
 &  & TL & 0.93 & 0.90 & 0.86 & 0.92 & 0.87 & 0.85 & 0.90 & 0.91\\
\cmidrule{3-11}
 &  & LA & \textbf{0.91} & \textbf{0.89} & \textbf{0.90} & \textbf{0.95} & \textbf{0.94} & \textbf{0.88} & \textbf{0.90} & \textbf{0.89}\\
\cmidrule{3-11}
 & \multirow{-3}{*}[1\dimexpr\aboverulesep+\belowrulesep+\cmidrulewidth]{\raggedright\arraybackslash Small} & NDM & 0.70 & 0.38 & 0.75 & \textbf{0.85} & 0.80 & 0.72 & 0.45 & 0.17\\
\cmidrule{2-11}
 &  & TL & 0.85 & 0.83 & 0.92 & 0.89 & 0.92 & 0.88 & 0.91 & 0.89\\
\cmidrule{3-11}
 &  & LA & \textbf{0.89} & \textbf{0.80} & \textbf{0.92} & \textbf{0.92} & \textbf{0.92} & \textbf{0.91} & \textbf{0.87} & \textbf{0.77}\\
\cmidrule{3-11}
 & \multirow{-3}{*}[1\dimexpr\aboverulesep+\belowrulesep+\cmidrulewidth]{\raggedright\arraybackslash Medium} & NDM & 0.56 & 0.20 & 0.75 & 0.84 & 0.86 & 0.83 & 0.25 & 0.06\\
\cmidrule{2-11}
 &  & TL & 0.90 & 0.90 & 0.92 & 0.93 & 0.94 & 0.89 & 0.90 & 0.92\\
\cmidrule{3-11}
 &  & LA & \textbf{0.93} & \textbf{0.76} & \textbf{0.98} & 1.00 & 0.95 & \textbf{0.97} & \textbf{0.76} & \textbf{0.42}\\
\cmidrule{3-11}
\multirow{-9}{*}[4\dimexpr\aboverulesep+\belowrulesep+\cmidrulewidth]{\raggedright\arraybackslash Low} & \multirow{-3}{*}[1\dimexpr\aboverulesep+\belowrulesep+\cmidrulewidth]{\raggedright\arraybackslash Large} & NDM & 0.70 & 0.08 & 0.79 & \textbf{0.85} & \textbf{0.87} & 0.82 & 0.12 & 0.02\\
\cmidrule{1-11}
 &  & TL & 0.92 & 0.91 & 0.88 & 0.94 & 0.92 & 0.92 & 0.91 & 0.92\\
\cmidrule{3-11}
 &  & LA & \textbf{0.91} & \textbf{0.89} & \textbf{0.88} & \textbf{0.95} & 0.94 & \textbf{0.93} & \textbf{0.92} & \textbf{0.75}\\
\cmidrule{3-11}
 & \multirow{-3}{*}[1\dimexpr\aboverulesep+\belowrulesep+\cmidrulewidth]{\raggedright\arraybackslash Small} & NDM & 0.43 & 0.36 & 0.64 & 0.83 & \textbf{0.92} & 0.79 & 0.36 & 0.23\\
\cmidrule{2-11}
 &  & TL & 0.90 & 0.92 & 0.89 & 0.94 & 0.97 & 0.93 & 0.91 & 0.90\\
\cmidrule{3-11}
 &  & LA & \textbf{0.85} & \textbf{0.84} & \textbf{0.90} & 0.98 & 0.98 & \textbf{0.95} & \textbf{0.83} & \textbf{0.21}\\
\cmidrule{3-11}
 & \multirow{-3}{*}[1\dimexpr\aboverulesep+\belowrulesep+\cmidrulewidth]{\raggedright\arraybackslash Medium} & NDM & 0.24 & 0.10 & 0.65 & \textbf{0.91} & \textbf{0.86} & 0.80 & 0.16 & 0.02\\
\cmidrule{2-11}
 &  & TL & 0.86 & 0.84 & 0.85 & 0.90 & 0.88 & 0.90 & 0.89 & 0.87\\
\cmidrule{3-11}
 &  & LA & \textbf{0.79} & \textbf{0.38} & \textbf{0.96} & 1.00 & 0.99 & 0.98 & \textbf{0.39} & 0.00\\
\cmidrule{3-11}
\multirow{-9}{*}[4\dimexpr\aboverulesep+\belowrulesep+\cmidrulewidth]{\raggedright\arraybackslash Medium} & \multirow{-3}{*}[1\dimexpr\aboverulesep+\belowrulesep+\cmidrulewidth]{\raggedright\arraybackslash Large} & NDM & 0.32 & 0.01 & 0.74 & \textbf{0.94} & \textbf{0.87} & \textbf{0.85} & 0.01 & 0.00\\
\cmidrule{1-11}
 &  & TL & 0.85 & 0.83 & 0.87 & 0.93 & 0.94 & 0.85 & 0.85 & 0.87\\
\cmidrule{3-11}
 &  & LA & \textbf{0.84} & \textbf{0.79} & \textbf{0.88} & \textbf{0.94} & \textbf{0.95} & \textbf{0.88} & \textbf{0.89} & \textbf{0.67}\\
\cmidrule{3-11}
 & \multirow{-3}{*}[1\dimexpr\aboverulesep+\belowrulesep+\cmidrulewidth]{\raggedright\arraybackslash Small} & NDM & 0.36 & 0.30 & 0.74 & 0.95 & 0.82 & 0.66 & 0.29 & 0.17\\
\cmidrule{2-11}
 &  & TL & 0.87 & 0.85 & 0.90 & 0.84 & 0.91 & 0.83 & 0.88 & 0.88\\
\cmidrule{3-11}
 &  & LA & \textbf{0.84} & \textbf{0.81} & \textbf{0.90} & 0.97 & 1.00 & \textbf{0.92} & \textbf{0.78} & 0.06\\
\cmidrule{3-11}
 & \multirow{-3}{*}[1\dimexpr\aboverulesep+\belowrulesep+\cmidrulewidth]{\raggedright\arraybackslash Medium} & NDM & 0.28 & 0.16 & 0.64 & \textbf{0.96} & \textbf{0.89} & 0.67 & 0.14 & \textbf{0.08}\\
\cmidrule{2-11}
 &  & TL & 0.86 & 0.81 & 0.83 & 0.92 & 0.97 & 0.88 & 0.86 & 0.92\\
\cmidrule{3-11}
 &  & LA & \textbf{0.89} & \textbf{0.37} & \textbf{0.97} & 1.00 & 1.00 & 1.00 & \textbf{0.40} & 0.00\\
\cmidrule{3-11}
\multirow{-9}{*}[4\dimexpr\aboverulesep+\belowrulesep+\cmidrulewidth]{\raggedright\arraybackslash High} & \multirow{-3}{*}[1\dimexpr\aboverulesep+\belowrulesep+\cmidrulewidth]{\raggedright\arraybackslash Large} & NDM & 0.28 & 0.03 & 0.82 & \textbf{0.99} & \textbf{0.96} & \textbf{0.83} & 0.02 & 0.00\\
\bottomrule
\end{tabular}
\end{table}

We note that the coverage rates for the true linkage model are
consistently around the 90\% nominal coverage rate, which serves as the
gold standard for the growth model performance. Comparing the LA and NDM
approaches, we see that the LA approach tends to outperform the NDM
approach and often by a substantial margin. In the instances where the
NDM coverage is closer to the nominal level, the coverage for the LA
approach is generally more conservative due to the uncertainty
propagation from the linkage stage of the modeling pipeline. These
results are in line with our expectations regarding the performance of
the different linkage approaches, and provide evidence that the
two-stage LA framework can reliably recover the parameters of interest
from a downstream model. In particular, the LA approach consistently has
better coverage for the growth asymptote parameter \(\beta_0\), which
may explain the differences that we observed in the estimates from the
empirical analysis shown in Figure \ref{coverage_plot}. We note that
coverage rates for \(\tau^2\) are generally lower than the nominal
level, most notably in the high noise scenarios for both the LA and NDM
approaches. This lower than nominal empirical coverage can be attributed
to the incorrect links that are introduced by both the LA and NDM
approaches as compared to the true linkage structure, which results in
an overestimation of the value of \(\tau^2\). However, even at higher
noise levels the empirical coverage for the covariate coefficients
remain at or above the nominal level for the LA approach. This suggests
that our interpretation of the impact of the covariates on growth is
robust to the error in the linkage. We also note that in the presence of
large amounts of noise, we may encounter identifiability issues with the
growth model parameters, as evidenced by the reduced coverage rates for
\(\beta_0\), \(\gamma\), and \(\tau^2\) in the LA and NDM models.
Coverage results for the growth model parameters for datasets with
\(\alpha = 2\) and \(\alpha = 3\) were similar and may be found in
Appendix D of the Supplementary Materials
(\cite{drewBayesianRecordLinkage2025sup}).

\section{Discussion and Future Work}\label{section7}

In this paper we have established a two-stage modeling framework built
around a record linkage model for spatial location data which serves as
the first step in a modeling pipeline constructed for multi-temporal
location data. We demonstrated the efficiency and scalability of this
approach for analyzing LiDAR derived individual tree characteristic data
and provided a general schematic for using the LA approach for two-stage
modeling to obtain equivalent inference for the downstream model
parameters to the marginal inference obtained from a joint model for the
linkage, latent spatial process, and downstream model. This framework
enables researchers to investigate growth trends as a function of
topographic information at a spatial scale that was previously difficult
to achieve and provides flexibility in examining a variety of downstream
models for different modeling objectives. It would also be straight
forward to extend for use in a Bayesian model averaging construction, as
introduced by \cite{rafteryBayesianModelAveraging1997}, when considering
a variety of candidate models for the same downstream modeling objective
in lieu of a model selection procedure, as we employed in this
application. Another natural extension of our record linkage model would
be to applications with more than two files, though some care would need
to be taken in specifying the prior for the linkage structure to ensure
the correct identification of clusters containing records from multiple
files. This extension could also be applied in the context of streaming
data where the linkage structure is updated as new data is collected as
discussed by \cite{taylorFastBayesianRecord2024}.

We applied this two-stage framework to investigate individual specific
growth-size curves of conifer species on Snodgrass Mountain in the
Southern Rocky Mountains of Colorado. We were able to quantify the
impact of several key topographic covariates that serve as proxies for
energy and water availability, which have been hypothesized to be
limiting factors on growth for conifer species in this region. We
demonstrated the effectiveness of our modeling approach in a series of
numerical experiments on simulated data, in the absence of a ground
truth dataset, and implemented a simulation framework for generating
data arising from a multi-temporal process as a function of the data
model from the linkage model and a general downstream growth model. This
approach provides researchers with an alternative tool for model testing
and validation in the absence of data with known linkage structure and
degrees of measurement error from the various processes involved.

As an alternative to the image alignment framework of
\cite{greenBayesianAlignmentUsing2006}, one could consider an additional
pre-processing step utilizing an image registration approach to
transform the observed point clouds. The image registration literature
is extensive, as discussed by \cite{zitovaImageRegistrationMethods2003},
and includes a variety of approaches that may be utilized in the context
of forestry data. For example, \cite{ferrazFusionNASAAirborne2018a}
introduced an approach for generating a fused high density vegetation
point cloud using a time series of low density LiDAR scans taken at
different times from the NASA-JPL Airborne Snow Observatory. Their
method similarly estimates a transformation matrix to align the point
clouds, although it uses an iterative procedure and relies on the use of
a collection of ``tie objects'' to estimate the transformation. However,
in the context of our application, in addition to misalignment, we
observe variability in the point clouds and derived digital surface
models such that the estimated tree crowns have different shapes and
sizes across scans due to the growth of the trees as seen in the inset
of Figure \ref{emp_rast_plot} panel c). As a result of the noise in the
LiDAR data and the biological processes of tree growth, we have limited
``tie objects'' available in our study domain with fixed shapes and
locations that would be usable in an image registration procedure.
Consequently, we believe that the image alignment framework of
\cite{greenBayesianAlignmentUsing2006}, which enables a fully Bayesian
implementation of the record linkage model that incorporates the
uncertainty inherent in the LiDAR scanning process when identifying
which records correspond to the same unobserved latent locations, is a
more appropriate choice.

While our modeling approach is flexible and scalable, it can be
sensitive to the specification of hyperparameters, like the maximum
number of unique individuals across datasets, in facilitating the
linkage. There is also a clear relationship with the amount of noise in
the observed spatial locations and the performance of the linkage model,
so care must be taken when considering the efficacy of a record linkage
approach with extremely noisy data. Our modeling approach attempts to
decompose the observed distortions in the data to more accurately
address the possible sources of error, but these mechanisms are somewhat
dependent on the spatial scale of the data (i.e.~systematic rotation in
a scan). We also note that while the LA approach gives researchers a
high degree of freedom, a joint modeling approach where the downstream
modeling objective influences the linkage will likely lead to improved
performance across the modeling pipeline if the downstream task can be
well specified. In our future work, we plan to explore the viability of
a joint modeling approach for similar problems which depend on
multi-temporal spatial location data.

\begin{acks}[Acknowledgments]
I.B. is also affiliated with the Clark Family School of Environment and Sustainability at Western Colorado University.

The authors would like to thank the Associate Editor, the Editor, and the referees for their insightful comments that markedly improved the
quality of this paper.
\end{acks}

\begin{funding}
A.K. was partially supported by NSF CAREER SES-2338428.

I.B. was partially supported by the Environmental System Science program, U.S. Department of Energy, Office of Biological and Environmental Research - DOE-DE-SC0023029.
\end{funding}

%%%%%%%%%%%%%%%%%%%%%%%%%%%%%%%%%%%%%%%%%%%%%%
%% Supplementary Material, including data   %%
%% sets and code, should be provided in     %%
%% {supplement} environment with title      %%
%% and short description. It cannot be      %%
%% available exclusively as external link.  %%
%% All Supplementary Material must be       %%
%% available to the reader on Project       %%
%% Euclid with the published article.       %%
%%%%%%%%%%%%%%%%%%%%%%%%%%%%%%%%%%%%%%%%%%%%%%
\begin{supplement}
\stitle{Web Supplement}
\sdescription{The web supplement contains an appendix with additional model specification and implementation details, a proof of Theorem 4.1, additional details for the empirical analysis, and additional simulation study details and results.}
\end{supplement}
\begin{supplement}
\stitle{Code Supplement}
\sdescription{All code used in the analysis along with step-by-step instructions for recreating the analysis, simulation study performed, and all figures and tables from the main paper and appendix. The code may also be found online in the following repository. (\url{https://github.com/lanedrew/SpRL})}
\end{supplement}
\begin{supplement}
\stitle{Data Supplement}
\sdescription{The empirical data used in the analysis can be obtained from the following repository. (\url{https://data.ess-dive.lbl.gov/datasets/ess-dive-32482a38131d613-20240503T212244619})}
\end{supplement}

%%  imsart-nameyear.bst  will be used to                   %%
%%  MR numbers will be added by VTeX.                      %%
%%                                                         %%
%%  Use \cite{...} to cite references in text.             %%
%%                                                         %%
%%%%%%%%%%%%%%%%%%%%%%%%%%%%%%%%%%%%%%%%%%%%%%%%%%%%%%%%%%%%%

%% if your bibliography is in bibtex format, uncomment commands:
\bibliographystyle{imsart-nameyear} % Style BST file
\bibliography{AOAS2061_references.bib}       % Bibliography file (usually '*.bib')
% \bibliography{'References\_Lib.bib'}  

%% or include bibliography directly:
% \begin{thebibliography}{}
% \bibitem[\protect\citeauthoryear{???}{???}]{b1}
% \end{thebibliography}

\end{document}